\documentclass[journal,transmag]{IEEEtran}

\usepackage{cite}
\usepackage{amsmath,amssymb,amsfonts}
\usepackage{graphicx}
\usepackage{multirow}
\usepackage{url}
\usepackage{overpic}
\usepackage{tikz}
\usetikzlibrary{arrows.meta,calc}
\newcommand{\panel}[3]{\put(#1,#2){\large\bfseries (#3)}}

\DeclareMathOperator{\sign}{sign}

\graphicspath{{./figures/}}

\begin{document}

\title{RAT-EC: Introducing a New Open-Source 3D Volume-Integral Eddy-Current Module}

\author{\IEEEauthorblockN{Daniel Mayr\IEEEauthorrefmark{1,}\IEEEauthorrefmark{2},
Jeroen van Nugteren\IEEEauthorrefmark{3},
Julien Dular\IEEEauthorrefmark{2},
Mariusz Wozniak\IEEEauthorrefmark{2},
Jasmin Smajic\IEEEauthorrefmark{1}, and
Juerg Leuthold\IEEEauthorrefmark{1}}
\IEEEauthorblockA{\IEEEauthorrefmark{1}Institute of Electromagnetic Fields (IEF), D-ITET, ETH Zurich, 8092 Zurich, Switzerland}
\IEEEauthorblockA{\IEEEauthorrefmark{2}Technology Department, CERN, 1211 Geneva, Switzerland}
\IEEEauthorblockA{\IEEEauthorrefmark{3}Little Beast Engineering, Budel, The Netherlands}%
\thanks{This work has been submitted to the IEEE for possible publication. Copyright may be transferred without notice, after which this version may no longer be accessible.
This work was partially funded by the High Field Magnet Programme at CERN.
Corresponding author: D. Mayr (e-mail: daniel.mayr@cern.ch).}}

\markboth{Submitted to IEEE Transactions on Magnetics}%
{Mayr \MakeLowercase{\textit{et al.}}: RAT-EC: A New Volume-Integral Eddy-Current Module}

\IEEEtitleabstractindextext{%
\begin{abstract}
We present RAT-EC, an open-source volume-integral eddy-current module for unstructured meshes, driven in the time domain by coils of arbitrary winding geometry. It is the first openly available code of this kind and part of the open-source RAT suite. Because only the conductor is meshed, neither the surrounding air nor an artificial outer boundary has to be modelled, which simplifies model preparation and brings problems with thin gaps and intricate air regions within reach of standard hardware. The formulation combines a semi-analytical evaluation of the singular and near-singular inductance terms, a multi-level fast multipole method that accelerates both the far-field inductance action and the coil field, topology-aware constraints for multiply connected conductors, and a matrix-free, GPU-accelerated time integration. The implementation is verified against closed-form solutions for a conducting sphere and a conducting ring, and is validated on the TEAM-7 benchmark. At 50\,Hz, RAT-EC reproduces the measured field as accurately as finite-element solutions. At 200\,Hz, it agrees with a converged finite-element model, but both deviate from published measurements, which suggests that this benchmark point should be re-examined. RAT-EC provides an open platform for volume-integral eddy-current simulation and method development.
\end{abstract}

}

\maketitle

\IEEEdisplaynontitleabstractindextext

\IEEEpeerreviewmaketitle

\section{Introduction}\label{sec:intro}
\IEEEPARstart{E}{ddy} currents are induced wherever a conductor is exposed to a time-varying magnetic field. Computing them in large three-dimensional (3D) structures remains one of the most demanding tasks in electromagnetic design. They dominate the losses in electrical machines and transformers, limit the switching speed of the gradient coils in magnetic-resonance scanners, and are the working principle of induction heating, non-destructive testing and wireless power transfer. In superconducting magnets, they shape the current and field transients during ramps and quenches. Reliable quench-protection simulation, the main bottleneck in the design of high-temperature superconducting (HTS) magnets, therefore requires accurate knowledge of the eddy currents induced in the surrounding conducting structures~\cite{wozniak2024fusillo,wozniak2025fusillo,taylor1979quench,eberhard1977quench}. Realistic models combine several complex coils with extended conducting volumes, making both the field computation and the eddy-current simulations expensive. A useful simulation tool must therefore be accurate, keep the modelling effort low (as model set-up consumes most of the engineering time), solve large problems efficiently, handle realistic 3D coils, model conductors with holes, allow thin gaps and multiply connected regions, and be openly available.

In practice, these problems are most often solved with finite-element (FE) methods, which resolve eddy currents in arbitrary 3D conductors and can cope with complex coil shapes~\cite{fiqus1,fiqus2}. Their cost, however, lies outside the conductor. The surrounding air has to be meshed, and the unbounded exterior has to be truncated by an artificial boundary. In addition, the thin insulation layers and narrow gaps of magnet windings are the most laborious part of the model to build. For the eddy-current problem, the volume-integral method (VIM) avoids these costs by meshing only the conducting domain and treating the free space analytically. Several efficient implementations have been reported. Rubinacci and co-workers accelerated 3D eddy-current computations with a fast multipole method, making large models tractable~\cite{rubinacci2004cariddi}. Bettini, Passarotto and Specogna developed a formulation on general polyhedral meshes~\cite{bettini2017vinco}, and the Udine group later placed the family of volume-integral eddy-current methods on a common rigorous foundation~\cite{passarotto2022foundations}. Meunier, Chadebec and Guichon in Grenoble solved the same eddy-current problem within an unstructured partial-element equivalent-circuit (PEEC) framework, representing the current density in the conductors with facet elements~\cite{nguyen2014facet,meunier2015facet}, and Torchio extended the approach from low to high frequencies with a volume PEEC formulation~\cite{torchio2019vpeec}. Equivalent-circuit and distributed-network models reduce the effort and the model complexity further, but they require the current paths to be assumed in advance, which reduces flexibility and accuracy on complex 3D geometries~\cite{venuturumilli}. Several of these methods are openly available as fast-Fourier-transform-accelerated codes on regular voxel grids~\cite{yucel2018voxhenry,guillod2025pypeec,torchio2022fftpeec}. The uniform cell size required by the transform, however, imposes a single resolution on the whole model and precludes local refinement of thin features. Related open codes gain speed by restricting the current path: the inductance extractor FastHenry discretizes conductors into current filaments~\cite{kamon1994fasthenry}, and thin-wall codes for the conducting structures of fusion devices confine the current to thin shells~\cite{hansen2025thincurr}. All of these openly available codes are formulated in the frequency domain or, when in the time domain, resolve no variation of the current through the conductor thickness. The time-varying field of the coils is often computed with a current line model (CLM), which represents each coil as a set of current-carrying line segments and evaluates the Biot--Savart law directly, giving a precise coil model without meshing.

An efficient, openly available tool for transient eddy-current problems on unstructured meshes, driven by windings of arbitrary shape, is still missing. This paper introduces RAT-EC, the eddy-current module of RAT (Rotation of vector potential A in Tesla), an open-source simulation suite~\cite{rat_group} used by research groups for superconducting magnet design and simulation~\cite{bagni2024quench,rossi2022hitriplus,maartensson2026ste1}. RAT-EC extends the suite from magnetostatic to magnetodynamic problems. It is written in parallelized and vectorized C++ and distributed as a Docker container that runs unchanged on a workstation or on a batch system such as HTCondor~\cite{htcondor_thain_2005}. RAT-EC combines established techniques: semi-analytical polyhedral integrals for the singular near and self terms~\cite{fabbri2008polyhedral,passarotto2022foundations}, a cohomology basis for conductors with holes~\cite{pellikka2013cohomology,dlotko2014lazy}, multipole acceleration of the far field~\cite{greengard1987fmm,cheng1999fmm}, and physics-based preconditioning of the Krylov iteration~\cite{ventre2022h2lu}. A single multipole engine evaluates both the impressed field of the winding, represented by a CLM, and the inductance operator, and both the far-field action and the sparse near-field product can be executed on a graphics processing unit (GPU). Since only the conductor is meshed, the module handles large problems with little modelling effort and moderate memory. Its application to a complete superconducting magnet is left to future work. Here, three verification and validation cases demonstrate that RAT-EC reproduces analytical references and the measurements of the TEAM-7 (Testing Electromagnetic Analysis Methods) benchmark~\cite{fujiwara1990team7} at the 50\,Hz operating point. At the 200\,Hz operating point, RAT-EC and an independently converged commercial finite-element model agree closely, but both deviate from measurements published in 1990, which indicates that this benchmark point should be re-examined. A reproducibility repository~\cite{rat_validation} regenerates every figure and table of this paper. RAT-EC is designed modularly: each numerical component is accessed through a defined interface and can be replaced without modifying the rest of the code, which makes RAT-EC a platform for method development.

Section~\ref{sec:method} presents the formulation and its implementation, Section~\ref{sec:verification} verifies RAT-EC against analytical solutions for a conducting sphere and a conducting ring, Section~\ref{sec:results} validates it on TEAM-7, and Section~\ref{sec:conclusion} concludes with an outlook.

\section{Methodology and Formulation}
\label{sec:method}

This section derives the governing equations of RAT-EC and describes their numerical treatment. Sections~\ref{sec:math-formulation}--\ref{sec:rhs} present the formulation and the discrete operators, Sections~\ref{subsec:clm} and~\ref{sec:L-matrix} the evaluation of the coil field and of the dense inductance operator, Sections~\ref{sec:gauging} and~\ref{sec:bc-cohom} the gauge and the boundary conditions for multiply connected conductors, and Section~\ref{sec:solver} the time integration and preconditioning.

\subsection{Mathematical Formulation}
\label{sec:math-formulation}

We start from Maxwell's equations in the magnetoquasistatic limit, in which the displacement current is neglected. Since $\nabla\!\cdot\!\mathbf{B}=0$, the magnetic flux density $\mathbf{B}$ derives from a magnetic vector potential $\mathbf{A}$, $\mathbf{B}=\nabla\times\mathbf{A}$, and Faraday's law for the electric field $\mathbf{E}$ becomes $\nabla\times(\mathbf{E}+\partial_t\mathbf{A})=\mathbf{0}$. Inside the conducting domain $\Omega$, Ohm's law reads $\mathbf{E}=\rho\mathbf{j}$, with $\rho$ the local resistivity and $\mathbf{j}$ the current density. The vector potential $\mathbf{A}$ can be separated into the impressed potential $\mathbf{a}_s$ of the source coils and the potential of the induced currents,
\begin{equation}\label{eq:apot}
\mathbf{A}=\mathbf{a}_s+\frac{\mu_0}{4\pi}\int_{\Omega}
\frac{\mathbf{j}(\mathbf{r}')}{\|\mathbf{r}-\mathbf{r}'\|}\,d\mathbf{r}' ,
\end{equation}
with $\mu_0$ the vacuum permeability. Together, these relations yield the electric field integral equation for eddy currents~\cite{albanese1985AT,passarotto2022foundations}. The second term in~\eqref{eq:apot} is a Biot--Savart integral over the conductor. The surrounding free space is described analytically by the Green function, so that only $\Omega$ has to be discretized and no artificial outer boundary is needed.

The current density must be solenoidal: in the magnetoquasistatic limit, $\nabla\!\cdot\!\mathbf{j}=0$ follows from Amp\`ere's law. The $\mathbf{T}$ formulation enforces this by construction, with $\mathbf{j}=\nabla\times\boldsymbol{\mathcal{T}}$ for an electric vector potential $\boldsymbol{\mathcal{T}}$, so that $\nabla\!\cdot\!\mathbf{j}=0$ holds identically rather than as a constraint~\cite{albanese1985AT,rubinacci2004cariddi,passarotto2022foundations}.

For the discretization, we expand the electric vector potential in first-order N\'ed\'elec/Whitney edge basis functions $\{\mathbf{w}_{e_i}\}$~\cite{nedelec1980mixed,bossavit1988whitney} on a tetrahedral mesh of $\Omega$, $\boldsymbol{\mathcal{T}}(\mathbf{r},t) = \sum_i T_i(t)\,\mathbf{w}_{e_i}(\mathbf{r})$, where edge $e_i$ carries the coefficient $T_i$. Throughout, $\boldsymbol{\mathcal{T}}$ denotes the continuous field and $\mathbf{T}$ the vector of its degrees of freedom (DOFs) $T_i$. By Stokes' theorem, the current $I_f$ through a face $f$ equals the circulation of $\boldsymbol{\mathcal{T}}$ along the face boundary $\partial f$, which under the edge-basis expansion reduces to a signed sum over the edges $e_j \subset \partial f$, weighted by the edge lengths $\ell_j$,
\begin{equation} \label{eq:Tsum}
I_f \;=\; \int_f \mathbf{j}\!\cdot\!\hat{\mathbf{n}}\,dS \;=\; \oint_{\partial f}\!\boldsymbol{\mathcal{T}}\!\cdot\!d\boldsymbol{\ell} \;=\; \sum_{e_j \subset \partial f} \pm\,T_j\,\ell_j,
\end{equation}
where $\hat{\mathbf{n}}$ is the unit face normal and the sign follows from the orientation of edge $e_j$ relative to face $f$. With the face currents collected in $\mathbf{I}$, \eqref{eq:Tsum} reads $\mathbf{I}=\mathbf{M}_{e\to f}\,\mathbf{T}$, where $\mathbf{M}_{e\to f}$ is the sparse edge-to-face incidence matrix. Solenoidality is inherited exactly, because the identity $\nabla\!\cdot\!(\nabla\times\,\cdot\,)=0$ carries over to the incidence structure of the mesh: for any $\mathbf{T}$, the signed face currents of every element sum to zero.

Testing the integral equation with the same edge basis (a Galerkin scheme whose test currents $\nabla\times\mathbf{w}_{e_i}$ are divergence-free by construction) yields the edge-indexed volume-integral system solved by RAT-EC, which requires no air mesh and preserves $\nabla\!\cdot\!\mathbf{j}=0$ exactly~\cite{albanese1985AT,rubinacci2004cariddi,passarotto2022foundations}:
\begin{equation} \label{eq:system}
\mathbf{R}\,\mathbf{T} + \mathbf{L}\,\frac{d}{dt}\mathbf{T} = -\,\frac{d}{dt}\mathbf{U},
\end{equation}
where $\mathbf{R}$, $\mathbf{L}$, and $\mathbf{U}$ are the resistance matrix, the inductance matrix, and the right-hand-side vector, respectively, with entries
\begin{align}
R_{ij} &= \int_{\Omega} \rho(\mathbf{r})\, \bigl(\nabla\times\mathbf{w}_{e_i}(\mathbf{r})\bigr)\!\cdot\! \bigl(\nabla\times\mathbf{w}_{e_j}(\mathbf{r})\bigr)\,d\mathbf{r}, \label{eq:Rij}\\[0.3em]
L_{ij} &= \frac{\mu_0}{4\pi} \int_{\Omega}\!\!\int_{\Omega} \frac{\bigl(\nabla\times\mathbf{w}_{e_i}(\mathbf{r})\bigr)\!\cdot\! \bigl(\nabla\times\mathbf{w}_{e_j}(\mathbf{r}')\bigr)}{\lVert \mathbf{r}-\mathbf{r}' \rVert}\, d\mathbf{r}\,d\mathbf{r}', \label{eq:Lij}\\[0.3em]
U_i &= \int_{\Omega} \mathbf{a}_s(\mathbf{r})\!\cdot\! \bigl(\nabla\times\mathbf{w}_{e_i}(\mathbf{r})\bigr)\,d\mathbf{r}. \label{eq:Ui}
\end{align}
Here $\mathbf{a}_s$ is the impressed vector potential of~\eqref{eq:apot}, supplied by the CLM of Section~\ref{subsec:clm}; it is not an unknown of the eddy-current problem. $\mathbf{R}$ is sparse, whereas $\mathbf{L}$ is dense because of the non-local kernel $1/\lVert\mathbf{r}-\mathbf{r}'\rVert$, and $-d\mathbf{U}/dt$ is the induced electromotive-force (EMF) vector driving~\eqref{eq:system}. The CARIDDI approach~\cite{rubinacci2004cariddi} and volume-PEEC~\cite{torchio2019vpeec} arrive at the same system with an equivalent parametrization of the solenoidal current space. Their algebraic equivalence to~\eqref{eq:system} is shown in~\cite{passarotto2022foundations}. Sections~\ref{sec:R-matrix} and~\ref{sec:rhs} describe the numerical evaluation of the operators (incidence maps, reconstruction of $\mathbf{j}$ at the quadrature points, and assembly of \eqref{eq:Rij}--\eqref{eq:Ui}), and Section~\ref{sec:L-matrix} the matrix-free treatment of $\mathbf{L}$.

\subsection{Discrete Operators and Resistance Matrix R}
\label{sec:R-matrix}

Two sparse incidence operators map between the DOF, face, and Gauss-point representations. The edge-to-face matrix $\mathbf{M}_{e\to f}$ is assembled from the signed edge--face incidences of~\eqref{eq:Tsum}, while the face-to-element matrix $\mathbf{M}_{f\to n}$ reconstructs the current density at the Gauss points of an element $v_h$ from its face fluxes $\{I_f\}$ with lowest-order Raviart--Thomas face basis functions $\mathbf{w}_f$~\cite{raviart1977mixed},
\begin{equation} \label{eq:RT}
\mathbf{j}(\mathbf{r}) \;=\; \sum_{f \subset \partial v_h} I_f\,\mathbf{w}_f(\mathbf{r}), \qquad \mathbf{r} \in v_h,
\end{equation}
evaluated at the quadrature points via the contravariant Piola transform~\cite{brezzi1991mixed}. The $\mathbf{w}_f$ do not introduce a second field description. Within each element, they compose exactly the curl of the edge basis functions (see~\eqref{eq:curl-expand}), so \eqref{eq:RT} evaluates the same current density $\mathbf{j}=\nabla\times\boldsymbol{\mathcal{T}}$ and serves only as an implementation convenience.

At the Gauss-point level, the resistance operator is diagonal. Each Gauss point $gp$ of element $v_h$ contributes the scalar weight
\begin{equation}\label{eq:Dgp}
d_{gp} = \rho_{gp}\,v_{gp}, \qquad v_{gp} = |J_{gp}|\,w_{gp},
\end{equation}
where $\rho_{gp}$ is the local resistivity, $|J_{gp}|$ the Jacobian determinant of the element map (a superscript, as in $J_{gp}^{h}$, identifies the element $v_h$ where several elements are involved), and $w_{gp}$ the quadrature weight on the reference element. Since $\mathbf{j}$ has three Cartesian components, $d_{gp}$ enters the diagonal matrix $D \in \mathbb{R}^{3N_{gp}\times 3N_{gp}}$ three times, where $N_{gp}$ is the total number of Gauss points. The incidence operators then yield the edge-to-edge resistance matrix by the congruence
\begin{equation}\label{eq:Re2e}
\mathbf{R} = \mathbf{M}_{e\to f}^T\,\mathbf{M}_{f\to n}^T\,D\,\mathbf{M}_{f\to n}\,\mathbf{M}_{e\to f}.
\end{equation}
$\mathbf{R}$ is sparse and symmetric, since only edges sharing an element interact. It is assembled once and reused throughout the time integration. We found a Gauss quadrature degree of 2 sufficient on tetrahedral elements. Without gauging, $\mathbf{R}$ is only positive \emph{semi}-definite. Every gradient field $\boldsymbol{\mathcal{T}} = \nabla\phi$ representable in the edge space produces no current, $\mathbf{j} = \nabla\times\nabla\phi = \mathbf{0}$, and hence $\mathbf{T}^{\top}\mathbf{R}\,\mathbf{T} = 0$ for its coefficient vector $\mathbf{T}$. The tree-cotree gauge (Section~\ref{sec:gauging}) removes these gradient modes and makes $\mathbf{R}$ positive definite.

\subsection{Right-Hand Side}
\label{sec:rhs}

To evaluate the right-hand side $-d\mathbf{U}/dt$ of~\eqref{eq:system}, the curl of each edge basis function in~\eqref{eq:Ui} is expressed by the Raviart--Thomas face basis functions around that edge~\cite{bossavit1988whitney},
\begin{equation}\label{eq:curl-expand}
\nabla\times\mathbf{w}_{e_i}(\mathbf{r}) = \sum_{f \in \mathcal{F}(e_i)} i(e_i,f)\,\mathbf{w}_f(\mathbf{r}),
\end{equation}
where $i(e_i,f) = \pm 1$ accounts for the orientation of face $f$ relative to edge $e_i$. Substitution into~\eqref{eq:Ui} and Gauss quadrature with the contravariant Piola transform~\cite{raviart1977mixed} give
\begin{equation}\label{eq:Uapprox}
U_i \;=\; \sum_{f \in \mathcal{F}(e_i)} i(e_i,f) \sum_{gp} w_{gp}\,|J_{gp}|\,\bigl(\mathbf{a}_s(\mathbf{x}_{gp})\cdot\mathbf{w}_f(\mathbf{x}_{gp})\bigr),
\end{equation}
with $w_{gp}$ and $|J_{gp}|$ as in~\eqref{eq:Dgp} and $\mathbf{w}_f(\mathbf{x}_{gp})$ the Piola-transformed face basis function at the Gauss point. Since $\mathbf{a}_s$ is the only time-dependent quantity, $dU_i/dt$ follows by replacing $\mathbf{a}_s$ with $d\mathbf{a}_s/dt$ in~\eqref{eq:Uapprox}. Section~\ref{subsec:clm} describes how $d\mathbf{a}_s/dt$ is obtained at the Gauss points.

All geometric factors in~\eqref{eq:Uapprox} (Piola-transformed basis functions, Jacobian determinants, quadrature weights, face--edge orientation signs, and the tree-cotree gauge filter) are constant during the simulation. They are therefore assembled once into a sparse operator $\mathbf{G}_\mathrm{rhs}$ of size $N_\mathrm{dof} \times 3N_{gp}$, with $N_\mathrm{dof}$ the number of edge DOFs retained after gauging, and every residual evaluation computes the right-hand side as a single sparse matrix-vector product,
\begin{equation}\label{eq:Grhs}
-\frac{d\mathbf{U}}{dt} = \mathbf{G}_\mathrm{rhs}\,\operatorname{vec}(d\mathbf{a}_s/dt),
\end{equation}
where $\operatorname{vec}(d\mathbf{a}_s/dt) \in \mathbb{R}^{3N_{gp}}$ stacks the Cartesian components of $d\mathbf{a}_s/dt$ at all Gauss points. The element geometry is thus not recomputed at each time step, and the right-hand side costs little compared with the dense inductance matrix-vector product. As for $\mathbf{R}$, a Gauss quadrature degree of 2 is sufficient.

The assembly loops over elements rather than edges, so that each face contribution in~\eqref{eq:Uapprox} is evaluated once and scattered with sign $i(e_i,f)$ into the rows of all edges bounding the face. The near and self blocks of the inductance matrix are assembled in the same way (Section~\ref{sec:L-near}).

\subsection{Accelerated Current Line Model}
\label{subsec:clm}

We compute the coil field with the \emph{Accelerated Current Line Model} (CLM) of the \texttt{rat-models} library~\cite{rat_group}.
Instead of discretizing the coil volume and the surrounding air, the CLM represents each coil by line currents. Every hexahedron of the coil mesh is replaced by one or several line currents between its two end faces.
Since the vector potential is linear in the current, the \emph{unit-current} vector potential and flux density are computed once at all Gauss points of the conductor mesh.
For a coil circuit $c$ carrying current $I_c$, the unit-current vector potential at a conductor Gauss point $\mathbf{r}$ is
\begin{equation}\label{eq:dAdI}
  \frac{\partial \mathbf{a}_c}{\partial I_c}(\mathbf{r})
  = \frac{\mu_0}{4\pi}
    \oint_{\Gamma_c} \frac{d\boldsymbol{\ell}'}{|\mathbf{r}-\mathbf{r}'|},
\end{equation}
where $\Gamma_c$ is the union of all line-current segments of circuit $c$.
A direct evaluation of~\eqref{eq:dAdI} at $N_{gp}$ Gauss points costs $\mathcal{O}(N_s N_{gp})$ operations, where $N_s$ is the number of line-current segments. With $N_s \sim 10^5$ for typical coils and $N_{gp} \sim 10^5$--$10^6$, this is prohibitive.
Our open-source C++ implementation of the Multi-Level Fast Multipole Method (MLFMM)~\cite{greengard1987fmm,kurzak2006fmm}, \texttt{rat-mlfmm}~\cite{rat_group}, reduces this cost to $\mathcal{O}(N_{gp} + N_s)$ by grouping the elements in an adaptive octree and compressing far-field interactions into multipole expansions of order $p$ (see Section~\ref{sec:L-far}).

Once precomputed, the contributions of all $N_c$ circuits are superimposed at every time step.
At time $t$, the rate of change of the impressed vector potential $\mathbf{a}_s$ at Gauss point $\mathbf{r}_i$ is
\begin{equation}\label{eq:dAdt}
  \frac{d\mathbf{a}_s}{dt}\bigg|_{\mathbf{r}_i}
  = \sum_{c=1}^{N_c}
    \left.\frac{\partial \mathbf{a}_c}{\partial I_c}\right|_{\mathbf{r}_i}
    \frac{dI_c}{dt}(t),
\end{equation}
where $dI_c/dt(t)$ is a user-supplied function of time and $\partial\mathbf{a}_c/\partial I_c$ is computed once at the start of the simulation. The induced electric field $\mathbf{E}_s = -\,d\mathbf{a}_s/dt$ drives the eddy currents.
Alternatively, a spatially uniform $d\mathbf{B}/dt$, e.g.\ from a large background solenoid, can be prescribed directly. The corresponding $d\mathbf{a}_s/dt$ is then known in closed form, and no MLFMM evaluation is required.
In both cases, $d\mathbf{a}_s/dt$ is passed as a $3N_{gp}$-vector to the operator $\mathbf{G}_{\mathrm{rhs}}$ of Section~\ref{sec:rhs}, so the source term costs one sparse matrix-vector product per time step.

Here, $I_c(t)$ is prescribed as an ideal current source, as in the TEAM-7 benchmark, and the back-EMF induced in the coil by the eddy currents is neglected. Where this reaction matters, most prominently for a magnet discharging through its own conducting support structure, $I_c(t)$ must become an unknown of the system. This two-way circuit coupling is left to future work.

\subsection{Inductance Matrix L}
\label{sec:L-matrix}

\subsubsection{Challenges of the Inductance Matrix}
\label{sec:L-dense-challenge}

The treatment of $\mathbf{L}$ is the main difficulty of the volume-integral method, as it poses three challenges that must be met simultaneously.
The first challenge is \emph{storage}. Since the kernel $1/|\mathbf{r}-\mathbf{r}'|$ in \eqref{eq:Lij} couples every element with every other element, $\mathbf{L}$ is dense, both in the edge representation of \eqref{eq:Lij} and in the face representation from which it is assembled (Section~\ref{sec:L-regimes}). The number of faces $F$ is of the same order as the number of edges. Storing all $\mathcal{O}(F^2)$ entries in double precision requires about $0.8$\,GB for $F=10^4$ and $8$\,TB for $F=10^6$, which is infeasible for realistic meshes~\cite{passarotto2022foundations}.
The second challenge is \emph{computational cost}. The product $\mathbf{L}\mathbf{v}$ is evaluated at every Krylov iteration (Section~\ref{sec:solver}), i.e.\ hundreds to thousands of times per simulation, and its direct $\mathcal{O}(F^2)$ evaluation is prohibitive already for moderate meshes. In~\cite{passarotto2022foundations}, a direct implementation required more than 24 hours per time step for $F=10^6$. This product dominates the cost of volume-integral solvers, and accelerating it is the central task of every VIM implementation~\cite{passarotto2022foundations,vacalebre2023lowrank}.
The third challenge is the \emph{kernel singularity}. The integrand diverges where source and target points coincide, as they do in all self-interaction integrals. Standard quadrature rules, which are exact only for polynomials up to their design degree, cannot capture this behaviour, so a dedicated treatment is required (Section~\ref{sec:L-near})~\cite{fabbri2008polyhedral,passarotto2022foundations}.

RAT-EC meets these challenges with a decomposition into self, near, and far interactions (Section~\ref{sec:L-regimes}), MLFMM acceleration of the far field (Section~\ref{sec:L-far}), and semi-analytical singularity extraction combined with the MAGICA factorization for the near and self terms (Sections~\ref{sec:L-near} and~\ref{sec:L-magica}).

\subsubsection{Three-Regime Decomposition: Self, Near, and Far}
\label{sec:L-regimes}

We partition all mesh element pairs into three regimes according to their spatial relationship.
\emph{Self} interactions couple an element with itself; the kernel of~\eqref{eq:Lij} is then singular, and standard quadrature cannot be applied.

\emph{Near} interactions couple elements in the MLFMM source-to-target interaction list, typically immediate octree neighbours, which are too close for an accurate multipole expansion. Their kernel is nearly singular. A topological criterion such as shared nodes is insufficient to identify them, since geometrically close elements of different mesh regions may share no node.
\emph{Far} interactions comprise all remaining, well-separated pairs. The kernel is smooth on the element scale and can be approximated to any accuracy by a multipole expansion of sufficient order, which in practice is limited by memory, runtime, and floating-point precision.
The near and self interactions are expensive to compute but sparse; they are computed once and stored as compressed sparse factors (Section~\ref{sec:L-near}). The far interactions, which make the operator dense, are evaluated by the MLFMM at every matrix-vector product without forming the matrix (Section~\ref{sec:L-far}).

Let $\mathcal{A}_{\mathrm{FMM}}(\hat{\mathbf{J}})$ denote the MLFMM operator that maps the effective current dipole moments $\hat{\mathbf{J}}$ [A\,m] of the elements to the far-field vector potential at the element barycenters. Since this operator is available only through its action, not as a matrix or an inverse, the linear systems must be solved iteratively and matrix-free, which motivates the inexact Newton--Krylov scheme of Section~\ref{sec:solver}.
With $\dot{\mathbf{T}}\equiv d\mathbf{T}/dt$, the matrix-vector product becomes
\begin{align}
\mathbf{L}\,\dot{\mathbf{T}} \;\approx\;&
\underbrace{%
  \mathbf{M}_{e\to f}^{\top}\,\mathbf{L}_{f}^{\mathrm{n+s}}\,\mathbf{M}_{e\to f}\,\dot{\mathbf{T}}
}_{\substack{\text{near \& self}\\[2pt]\text{precomputed sparse factors}}}\notag\\
&+\;
\underbrace{%
  \mathbf{M}_{e\to f}^{\top}\,\mathbf{M}_{f\to n}^{\top}\,\mathrm{diag}(\mathbf{v}_{gp})\;
  \mathcal{A}_{\mathrm{FMM}}\!\bigl(\hat{\mathbf{J}}\bigr)
}_{\substack{\text{far field}\\[2pt]\text{MLFMM, on-the-fly}}},
\label{eq:L-mvp}\\[2pt]
&\hat{\mathbf{J}} \;=\; \mathbf{V}\,\mathbf{M}_{f\to n}\,\mathbf{M}_{e\to f}\,\dot{\mathbf{T}},\notag
\end{align}
where $\mathbf{L}_{f}^{\mathrm{n+s}}$ is the precomputed sparse face-space inductance of the near and self interactions, $\mathbf{v}_{gp}$ stacks the quadrature volumes $v_{gp}$ of~\eqref{eq:Dgp}, and the block-diagonal matrix $\mathbf{V}$ integrates the Gauss-point current densities [A/m$^2$] over each element to dipole moments $\hat{\mathbf{J}}$ [A\,m].

\subsubsection{Far-Field Interactions via MLFMM}
\label{sec:L-far}

For well-separated pairs, each source element in~\eqref{eq:Lij} is replaced by a point source at its barycenter that carries the volume-integrated element current.
The face fluxes $\mathbf{M}_{e\to f}\,\dot{\mathbf{T}}$ are mapped to Gauss-point current densities by $\mathbf{M}_{f\to n}$ and integrated over each element by $\mathbf{V}$, which gives the dipole moments $\hat{\mathbf{J}}$.
$\mathcal{A}_{\mathrm{FMM}}$ treats them as current dipoles at the element barycenters and returns their far-field vector potential,
which $\mathbf{M}_{f\to n}^{\top}\,\mathrm{diag}(\mathbf{v}_{gp})$ projects back to face space as in \eqref{eq:L-mvp}.
Between octree leaf clusters that the MLFMM admissibility criterion classifies as far, $\mathcal{A}_{\mathrm{FMM}}$ uses multipole expansions of order $p$~\cite{greengard1987fmm,cheng1999fmm}. An optional extended interaction list in \texttt{rat-mlfmm} also assigns next-nearest neighbours to the near regime. At $p=8$, we observe relative errors of the vector potential of ${\sim}2\times 10^{-5}$.
The same \texttt{rat-mlfmm} library as for the CLM (Section~\ref{subsec:clm}) performs this evaluation at every Krylov iteration.

\begin{figure}[!tbp]
  \centering
  \includegraphics[width=\columnwidth]{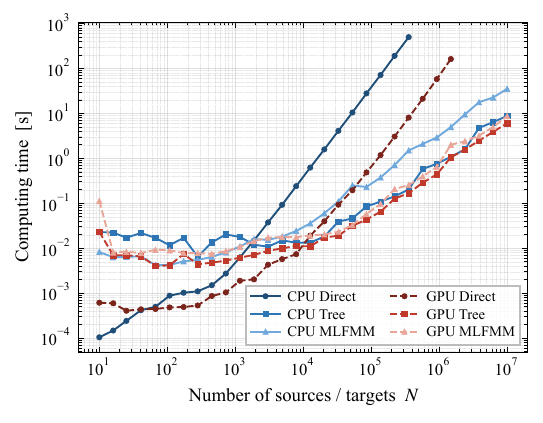}\par
  \vspace{-7pt}
  \caption{Wall-clock time per evaluation of the open-source MLFMM implementation \texttt{rat-mlfmm} versus the number of sources and targets for the direct sum, the MLFMM, and the one-off tree setup, on CPU (solid) and GPU (dashed). The MLFMM scales linearly instead of quadratically, which keeps the field evaluations of every Krylov iteration affordable at problem sizes where the direct sum is intractable.}
  \label{fig:fmm-scaling}
\end{figure}

\texttt{rat-mlfmm} provides CPU and GPU backends. For large problems, the GPU is about an order of magnitude faster, most visibly in the \emph{direct} source-to-target step, which RAT-EC replaces by the precomputed $\mathbf{L}_{f}^{\mathrm{n+s}}$. Fig.~\ref{fig:fmm-scaling} shows wall-clock times on a Threadripper~7980X workstation with an RTX~A5000 GPU for equal numbers of sources and targets, randomly distributed in a cube, at expansion order $p = 5$. The MLFMM curves include the direct source-to-target step.
The direct sums scale as $\mathcal{O}(N^2)$. Without the MLFMM overhead, they are faster below $N \approx 10^3$--$10^4$, but they exceed $10^2$~s at $N \approx 10^5$ on the CPU and at $N \approx 10^6$ on the GPU. The MLFMM scales linearly, with a slope of 1:1 in the log--log plot compared with 2:1 for the direct sums, and is faster for larger problems. The octree setup, plotted separately as \emph{Tree}, is a one-off cost. Since the conductor mesh does not move, the tree is built once and reused for all matrix-vector products.

\subsubsection{Near and Self Interactions}
\label{sec:L-near}

The near and self integrals cannot be evaluated by the MLFMM, since the point-source approximation fails at short range. Their accuracy is critical. The near+self block contains the singular and near-singular part of the kernel, and an inaccurate evaluation degrades the conditioning of the linear system and the stability of the time integration~\cite{passarotto2022foundations,fabbri2008polyhedral}.
For each element pair $(v_h, v_k)$, the contribution to $L_{ij}$ is evaluated directly. Substituting~\eqref{eq:curl-expand} into both curl factors of~\eqref{eq:Lij} and applying Gauss quadrature with the contravariant Piola transform to both volume integrals gives
\begin{equation} \label{eq:Lapprox}
  L_{ij}
  = \frac{\mu_0}{4\pi}
    \sum_{gp=1}^{G_h}\sum_{gp'=1}^{G_k}
    \frac{s_i \cdot s_j}{\lvert \mathbf{x}_{gp}^{h}-\mathbf{x}_{gp'}^{k}\rvert}
    \lvert \det J_{gp}^{h} \rvert
    \lvert \det J_{gp'}^{k} \rvert
    \, w_{gp}^{h}\, w_{gp'}^{k}
  \end{equation}
  with
  \begin{align*}
  s_i &= \sum_{f\subset\partial v_h} i(e_i,f)\;
         \mathbf{w}_f^{v_h}\!\big(\mathbf{x}_{gp}^{h}\big),\\
  s_j &= \sum_{f\subset\partial v_k} i(e_j,f)\;
         \mathbf{w}_f^{v_k}\!\big(\mathbf{x}_{gp'}^{k}\big),
  \end{align*}
where $G_h$ and $G_k$ are the number of Gauss points in $v_h$ and $v_k$; $\mathbf{x}_{gp}^{h}$, $\mathbf{x}_{gp'}^{k}$ their position vectors; $|\det J_{gp}^h|$, $|\det J_{gp'}^k|$ the Jacobian determinants of the element maps; $w_{gp}^h$, $w_{gp'}^k$ the reference quadrature weights; and $\mathbf{w}_f^{v_h}$, $\mathbf{w}_{f'}^{v_k}$ the Piola-transformed Raviart--Thomas face basis functions~\cite{raviart1977mixed}.
The vectors $s_i$ and $s_j$ are the values of $\nabla\times\mathbf{w}_{e_i}$ and $\nabla\times\mathbf{w}_{e_j}$ at the Gauss points of $v_h$ and $v_k$; their dot product accounts for the relative current directions.

In the near regime ($h \neq k$), the denominator in~\eqref{eq:Lapprox} never vanishes, since Gauss points of different elements cannot coincide.
In the self regime ($h = k$), the denominator vanishes for $gp = gp'$, and~\eqref{eq:Lapprox} diverges regardless of the number of quadrature points. Standard Gauss quadrature without singularity treatment produces errors of up to 200\,\% on the diagonal~\cite{passarotto2022foundations}. A simple remedy replaces each singular term $gp = gp'$ by the analytical self-inductance of a sphere with the Gauss-point volume $|v_{gp}|$ of~\eqref{eq:Dgp}, derived with a symbolic solver,
\begin{equation} \label{eq:Lsphere}
  L_s
  = \sign  w_{gp}   \, \frac{\mu_0}{2}   \, s_i \cdot s_i
  \left (
    \frac{3\, \lvert \det J_{gp} \rvert \, \lvert w_{gp} \rvert }{4\pi} \right )^\frac{2}{3}
    \,
    \lvert \det J_{gp} \rvert
    \, w_{gp}
  \end{equation}
We refer to Gauss quadrature with this analytical-sphere diagonal replacement as \emph{optimized Gauss}.

\begin{figure}[!t]
  \centering
  \includegraphics[width=\columnwidth]{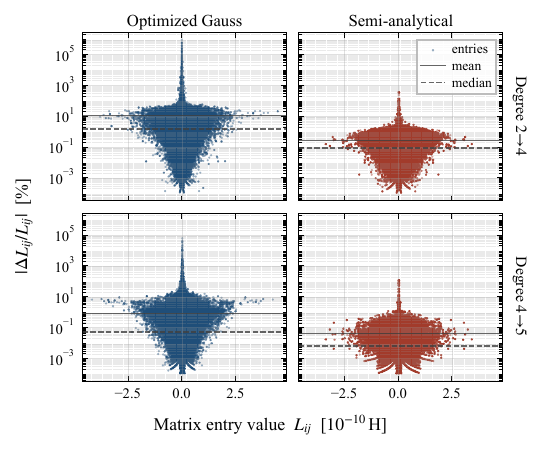}\par
  \vspace{-7pt}
  \caption{Convergence of the near-interaction entries $L_{ij}$ with Gauss degree: relative change $|\Delta L_{ij}/L_{ij}|$ for the degree steps $2\!\to\!4$ (top) and $4\!\to\!5$ (bottom), with optimized Gauss quadrature (left, blue) and with RAT-EC's semi-analytical scheme (right, red). At equal degree, the semi-analytical scheme converges about one order of magnitude tighter.}
  \label{fig:L-convergence-near}
\end{figure}

\begin{figure}[!t]
  \centering
  \includegraphics[width=\columnwidth]{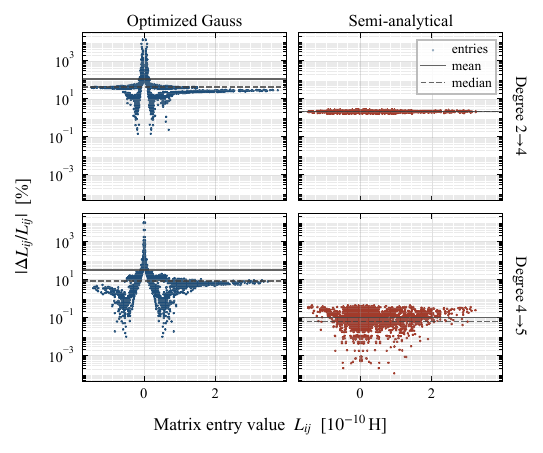}\par
  \vspace{-7pt}
  \caption{Convergence of the self-interaction entries $L_{ij}$ with Gauss degree: relative change $|\Delta L_{ij}/L_{ij}|$ for the degree steps $2\!\to\!4$ (top) and $4\!\to\!5$ (bottom), with optimized Gauss quadrature (left, blue) and with RAT-EC's semi-analytical scheme (right, red). At equal degree, the semi-analytical scheme converges one to two orders of magnitude tighter.}
  \label{fig:L-convergence-self}
\end{figure}

This removes the singularity and yields a convergent result, but the remaining double sum converges too slowly for realistic meshes. We quantified this on a sphere mesh of ${\sim}10^4$ tetrahedra with ${>}10^6$ near and ${\sim}10^4$ self element pairs. The left (blue) columns of Figs.~\ref{fig:L-convergence-near} and~\ref{fig:L-convergence-self} show the relative change $|\Delta L_{ij}/L_{ij}|$ of each entry. Degree~3 was excluded, since its large, negative weight destabilizes the near-singular integrand~\cite{keast1986,shunn2012}.

For the near regime, raising the degree $2\!\to\!4$ changes the entries by $11\%$ on average (median $1.5\%$), and $4\!\to\!5$ by $0.8\%$ (median $0.05\%$). Degree~5 would thus be accurate enough, but its assembly time (Table~\ref{tab:L-near-self-timings}) is prohibitive for the targeted mesh sizes, despite parallelization and the element-ordered assembly of Section~\ref{sec:rhs}.

The self regime behaves considerably worse: the step $2\!\to\!4$ leaves a mean change of $111\%$ (median $43\%$), and $4\!\to\!5$ still $31\%$ (median $8\%$), with isolated relative changes of $10^6$ for entries close to zero. The convergence is also non-monotonic, as the sphere correction~\eqref{eq:Lsphere} smooths the residual singularity but does not remove it.

Optimized Gauss quadrature is thus insufficient for the self integrals and only marginally adequate for the near integrals, which calls for a dedicated treatment of the kernel.

Unlike the Raviart--Thomas-weighted integrand $s_i\cdot s_j/|\mathbf{r}-\mathbf{r}'|$ of~\eqref{eq:Lapprox}, the kernel $\int_{v_k}1/|\mathbf{r}-\mathbf{r}'|\,d\mathbf{r}'$ over a polyhedral source of \emph{uniform} density has a simple closed form~\cite{fabbri2008polyhedral}. We therefore evaluate the integral with volume-uniform (VU) basis functions, which yield exactly the same inductance matrix as the Raviart--Thomas basis~\cite{passarotto2022foundations}\footnote{A lowest-order Raviart--Thomas current on a tetrahedron is $\mathbf{j}=\mathbf{a}+b\,\mathbf{r}$ for a constant vector $\mathbf{a}$ and scalar $b$, with constant divergence $\nabla\!\cdot\!\mathbf{j}=3b$. The solenoidal constraint $\mathbf{j}=\nabla\times\boldsymbol{\mathcal{T}}$ forces $b=0$ and collapses it to the constant field $\mathbf{a}$ that the VU basis spans. As $\mathbf{L}=\mathbf{M}_{e\to f}^{\top}\mathbf{L}_f\mathbf{M}_{e\to f}$ applies this projection to both the trial and test currents, the two bases yield identical inductance integrals.}, so that Fabbri's closed form applies without breaking the Whitney-form structure of the system.

The resulting scheme is \emph{semi-analytical}. The inner integral over the source element $v_k$ is evaluated analytically with~\cite[Eqs.~(17)--(21)]{fabbri2008polyhedral}, which gives a smooth, finite field at every target point, including points inside $v_k$. The outer integral over the target element $v_h$ is then computed by standard Gauss quadrature of this smooth field, so the singularity never enters the numerical integration.

Figs.~\ref{fig:L-convergence-near} and~\ref{fig:L-convergence-self} quantify the gain on the same sphere mesh. At equal Gauss degree, the changes of the semi-analytical scheme (right columns) are one to two orders of magnitude smaller than those of optimized Gauss quadrature (left columns) in both blocks. For the step $4\!\to\!5$, the mean relative change falls to $0.04\,\%$ for the near and below $0.1\,\%$ for the self pairs, while the assembly is two to three times faster (Table~\ref{tab:L-near-self-timings}). Both degrees default to five. Lower degrees reduce the assembly time with some loss of accuracy, and the convergence plots guide this trade-off.

Since convergence alone does not exclude a bias, we also compared the computed self-inductance of a unit tetrahedron with a fully analytical value obtained with a symbolic solver. The deviation was below 0.1\,\%, which corroborates convergence to the correct value.

\begin{table}[!tbp]
\centering
\caption{One-Off Assembly Time of the Near and Self Blocks of $\mathbf{L}_f^{\mathrm{n+s}}$ With Optimized Gauss Quadrature and the Semi-Analytical Scheme, for Sphere Meshes of $V$ Tetrahedra and Gauss Degree $d$ (Intel 11th-Generation i7 CPU, 32~GB RAM, CPU Only)}
\label{tab:L-near-self-timings}
\setlength{\tabcolsep}{4pt}
\begin{tabular}{rcrrrrrrr}
\hline
         &     & \multicolumn{3}{c}{Semi-Analytical [s]} & \multicolumn{3}{c}{Optimized Gauss [s]} & \\
   $V$   & $d$ & near & self & total & near & self & total & $t_{\mathrm{OG}}/t_{\mathrm{SA}}$ \\
\hline
 1\,280 & 4 &   9.9 & 0.3 &  10.2 &   18.8 &  0.3 &   19.2 & $1.9\times$ \\
 1\,280 & 5 &  11.7 & 0.4 &  12.1 &   32.5 &  0.7 &   33.1 & $2.7\times$ \\
10\,240 & 4 & 766   & 1.8 & 768   & 1\,371 & 2.4  & 1\,373 & $1.8\times$ \\
10\,240 & 5 & 936   & 2.2 & 938   & 2\,508 & 4.2  & 2\,512 & $2.7\times$ \\
\hline
\end{tabular}
\end{table}

\subsubsection{MAGICA Factorization and Memory Optimization}
\label{sec:L-magica}

Evaluating~\eqref{eq:Lapprox} for all near and self pairs yields the sparse face-face inductance $\mathbf{L}_f^{\mathrm{n+s}}$ of~\eqref{eq:L-mvp}. Storing it entry by entry is redundant, since the contributions of all four faces of a tetrahedron share the same element integral. The MAGICA factorization~\cite{passarotto2022foundations} exploits this redundancy through the structure of the VU basis functions.
For each element $v_h$ and each of its faces $f_j$, the VU basis vector is
\begin{equation}\label{eq:VU-basis}
\tilde{\mathbf{e}}^{v_h}_{f_j}
= \frac{\mathbf{b}_{f_j} - \tilde{\mathbf{n}}_{v_h}}{|v_h|},
\end{equation}
where $\mathbf{b}_{f_j}$ is the face barycenter and $\tilde{\mathbf{n}}_{v_h}$ the dual node (barycenter) of $v_h$.
Collecting the Cartesian components of these vectors into three sparse $V\times F$ matrices $\tilde{\mathbf{E}}_x$, $\tilde{\mathbf{E}}_y$, $\tilde{\mathbf{E}}_z$, and defining the $V\times V$ scalar kernel matrix $\mathbf{N}$ with entries
\begin{equation}\label{eq:N-kernel}
N_{hk} = \frac{\mu_0}{4\pi}\,t^{hk},
\qquad
t^{hk} = \int_{v_h}\!\int_{v_k}\frac{d\mathbf{r}\,d\mathbf{r}'}{\lVert\mathbf{r}-\mathbf{r}'\rVert},
\end{equation}
the near and self inductance factors exactly as~\cite{passarotto2022foundations}
\begin{equation}\label{eq:MAGICA}
\mathbf{L}_f^{\mathrm{n+s}} = \tilde{\mathbf{E}}_x^T\,\mathbf{N}\,\tilde{\mathbf{E}}_x
           + \tilde{\mathbf{E}}_y^T\,\mathbf{N}\,\tilde{\mathbf{E}}_y
           + \tilde{\mathbf{E}}_z^T\,\mathbf{N}\,\tilde{\mathbf{E}}_z.
\end{equation}
The double integral $t^{hk}$ is exactly what the semi-analytical scheme of Section~\ref{sec:L-near} computes, with Fabbri's closed form for the inner and Gauss quadrature for the outer integral.
Thus, $\mathbf{L}_f^{\mathrm{n+s}}$ is never assembled. Only the sparse factors $\mathbf{N}$ and $\tilde{\mathbf{E}}_{x,y,z}$ are stored, which reduces the memory footprint on tetrahedral meshes by a factor of 3 to 5, depending on the geometry. The three chains of sparse matrix-vector products do not slow down the near+self term of~\eqref{eq:L-mvp}. Except on very small meshes (up to two times slower below ${\sim}10^2$ elements), the additional FLOPs are outweighed by the reduced memory traffic, so the factored form is faster overall.

The factors are assembled once per mesh and cached on disk for repeated runs.

\subsection{Gauging}
\label{sec:gauging}

The edge-space system~\eqref{eq:system} is \emph{under-determined}. Each face current $I_f = (\mathbf{M}_{e\to f}\,\mathbf{T})_f$ is a signed sum over its edges, so different edge potentials can produce the same face currents. The redundant component is the null space of $\mathbf{M}_{e\to f}$, i.e.\ the discrete gradients on the edge graph~\cite{passarotto2022foundations,ren1996rhs}. $\mathbf{R}$ and $\mathbf{L} = \mathbf{M}_{e\to f}^{\top}\mathbf{L}_f\mathbf{M}_{e\to f}$ inherit this null space and are singular. As a consequence, the preconditioner of Section~\ref{sec:solver} cannot be factored directly, and $\mathbf{T}$ can drift in the null space without physical effect until it corrupts the weighted RMS error norm that the time integrator uses to select its steps.

The standard remedy is a \emph{tree--cotree gauge}~\cite{passarotto2022foundations,albanese1985AT}. The edges of a spanning tree of the edge graph are set to $T_e = 0$, and only the cotree edges remain unknown. On this reduced space, $\mathbf{R}$ and $\mathbf{L}$ are symmetric positive definite, and the preconditioner is well defined. We build the tree once per mesh with the \texttt{SpanningTree} plugin of the Gmsh~\cite{Gmsh_ijnme} C++ API, with the conductor volume as the primary physical group and its boundary surface as an included subdomain.

The boundary condition introduces a second, topological constraint. Since the conductor is insulated by the surrounding air, the normal current on $\partial\Omega$ vanishes, i.e.\ $I_f = 0$ on every boundary face, which is enforced by $T_e = 0$ on every edge of $\partial\Omega$~\cite{passarotto2022foundations}. These edges are fixed before the tree is built. This suffices for simply connected conductors, whereas multiply connected conductors require a few additional cohomology DOFs (Section~\ref{sec:bc-cohom}).

We apply both eliminations by removing the corresponding columns of $\mathbf{M}_{e\to f}$, so that all derived operators inherit the reduction. On typical tetrahedral meshes, this reduces the number of edge DOFs, and hence the size of all matrices, by up to about $50\%$.

Alternatively, the tree elimination can be skipped and the full, singular edge system integrated. In the VIM literature, this is done in the high-frequency regime, where an ungauged system with a complex diagonal preconditioner outperforms the gauged one~\cite{passarotto2022foundations}. RAT-EC supports both variants through a runtime switch. The surface condition above is a boundary condition, not a gauge, and is always applied. Since the right-hand side is orthogonal to the gradient null space, the singular system is \emph{consistent}, and the generalized minimal residual (GMRES) method converges in the quotient space~\cite{brown1997gmres}.\footnote{With the null space unconstrained, the individual edge DOFs grow without bound (in practice exponentially) and only their superpositions (the face currents) converge to finite values, so a long integration eventually exhausts the finite-precision representation. This can be solved by keeping the tree--cotree structures of the gauged path at hand and, once the DOF magnitudes cross a threshold, projecting the state to the gauged representation and back, zeroing exactly the tree-edge components the gauge would have eliminated.} Only the preconditioner inherits the rank deficiency. RAT-EC factors it with corrections that modify only the preconditioner and leave the physics unchanged: a mid-spectrum diagonal shift~\cite{manteuffel1980}, a grad--div augmentation~\cite{biro1989}, a deflation projection~\cite{nicolaides1987deflation,brown1997gmres}, or a null-pivot-aware $LDL^{\!\top}$ factorization (MUMPS~\cite{mumps2001}). In low-frequency tests, both formulations perform equally within the measurement noise. Since the tree--cotree gauge always yields a regular system and is robust across problem sizes, it is the default in RAT-EC. A systematic study of the ungauged variant is left to future work.

\subsection{Boundary Conditions and Multiply Connected Domains}
\label{sec:bc-cohom}

\begin{figure*}[!tbp]
  \centering
  \includegraphics[width=0.8\textwidth]{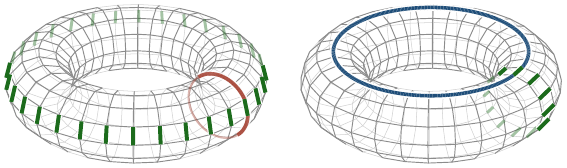}\par
  \vspace{-3pt}
  \caption{Cohomology generators on a structured surface mesh of a torus. \emph{Left:} poloidal homology cycle (red) and associated generator of $H^1(\partial\Omega;\mathbb{Z})$ (green edges). \emph{Right:} toroidal cycle (blue) and associated generator (green). Pinning every surface edge would suppress the current around the hole. One scalar unknown on the green edges per hole restores this current exactly while no current crosses the boundary.}
  \label{fig:torus-cohom}
\end{figure*}

The condition $T_e = 0$ on all edges of $\partial\Omega$ suffices for simply connected conductors but over-constrains geometries with holes. By Stokes' theorem~\eqref{eq:Tsum}, the net current through a surface equals the sum of the edge potentials around its boundary. On a torus, the red loop in the left panel of Fig.~\ref{fig:torus-cohom} is a poloidal cycle on $\partial\Omega$ whose edges are all fixed to zero, so the net current through the disk it spans vanishes and the current circulating \emph{around the hole} is suppressed.

The established remedy is based on cohomology~\cite{passarotto2022foundations,pellikka2013cohomology,dlotko2014lazy} with a few additional DOFs introduced exactly where the surface condition would block a physical current. A set $\mathcal{H}$ of surface edges (green in the left panel of Fig.~\ref{fig:torus-cohom}) is chosen that intersects every such poloidal loop without opening a current path through a boundary face. Two conditions define $\mathcal{H}$: no closed cycle on $\partial\Omega$ may avoid $\mathcal{H}$, so that the remaining surface edges no longer suppress the net current around the hole; and every boundary face must share either zero or exactly two edges with $\mathcal{H}$. A single scalar coefficient $t_\mathcal{H}$, attached to these edges with alternating signs, then cancels in the Stokes sum~\eqref{eq:Tsum} of every boundary face, which preserves $I_f = 0$ on the boundary while allowing a net current around the hole. In topological terms, $\mathcal{H}$ is a generator of the first cohomology group $H^1(\partial\Omega;\mathbb{Z})$ of the boundary surface, and the number of required generators equals the first Betti number $b_1(\Omega)$ of the conductor, i.e.\ one per through-hole~\cite{passarotto2022foundations,pellikka2013cohomology}.

We compute the generators with the cohomology solver of Gmsh~\cite{Gmsh_ijnme,pellikka2013cohomology} through its C++ API, applied to the two-dimensional boundary mesh of the conductor. For each hole, it returns two generators, the \emph{poloidal} and the \emph{toroidal} one sketched in Fig.~\ref{fig:torus-cohom}, of which only one is required since $b_1(\Omega)=1$. By default, and for all results in this paper, the retained generator is selected manually for each geometry. Two alternatives avoid this manual choice. The first, the ``lazy'' cohomology of D{\l}otko and Specogna~\cite{dlotko2014lazy}, keeps both generators. The superfluous one adds an unobservable DOF per hole and reintroduces a rank deficiency of the kind discussed in Section~\ref{sec:gauging}. RAT-EC handles it with the same preconditioner corrections, restricted to the few cohomology diagonal entries, and the physics remains correct within run-to-run noise. With adaptive time stepping, however, the step-size controller follows the undetermined null direction, which increased the number of accepted steps on the ring benchmark by a factor of $1.2$--$2.8$. As a second alternative, we propose an algebraic test that identifies the superfluous generator automatically. A candidate generator with signed edge vector $\mathbf{h}$ is superfluous if and only if its face-current pattern can already be represented by the gauged interior DOFs, i.e.\ if $\mathbf{M}_{e\to f}\,\mathbf{h} \in \operatorname{range}(\mathbf{W})$, where $\mathbf{W} = \mathbf{M}_{e\to f}\,\mathbf{P}$ is the gauged interior map and $\mathbf{P}$ the interior-DOF projection introduced below. The test requires one sparse Cholesky factorization of $\mathbf{W}^{\!\top}\mathbf{W}$ and one triangular solve per candidate. On the geometries tested so far, it separates the required from the superfluous generator by fourteen orders of magnitude in the relative residual at negligible cost (${\sim}0.1$\,s for the ring). Its systematic validation is left to future work.

In the implementation, the unknowns are represented on two levels. The surface and tree edges are removed by deleting the corresponding columns of $\mathbf{M}_{e\to f}$, but the cohomology edges are \emph{kept} in the edge representation, so that $\mathbf{R}\mathbf{T}$, $\mathbf{L}\dot{\mathbf{T}}$ and $d\mathbf{U}/dt$, which all act on edge potentials, are evaluated correctly on the cycle. The results are then reduced to DOF space. Two sparse matrices connect the spaces: the projection $\mathbf{P}$, which maps each interior cotree DOF to its kept edge, and the cohomology matrix $\mathbf{H}$, whose columns hold the alternating $\pm 1$ pattern of each retained generator (the entries are $\pm 1$ rather than inverse edge lengths because of the scaling in Section~\ref{sec:solver}). With one scalar DOF per retained generator collected in $\mathbf{t}_\mathcal{H} \in \mathbb{R}^{N_\mathcal{H}}$, $N_\mathcal{H} = b_1(\Omega)$, the kept-edge state is $\mathbf{T}_\mathrm{kept} = \mathbf{P}\,\mathbf{T}_\mathrm{int} + \mathbf{H}\,\mathbf{t}_\mathcal{H} = \mathbf{P}_\mathrm{cat}\,\mathbf{y}$, where $\mathbf{P}_\mathrm{cat} = [\,\mathbf{P}\;\;\mathbf{H}\,]$ is cached as the change of basis from the DOF vector $\mathbf{y} = (\mathbf{T}_\mathrm{int}^{\!\top},\,\mathbf{t}_\mathcal{H}^{\!\top})^{\!\top}$~\cite{bettini2017vinco,passarotto2022foundations}. The construction adds little computational cost. Unlike generic algebraic null-space methods, whose dense factors grow at least quadratically with the mesh size, the Gmsh cohomology solver~\cite{pellikka2013cohomology} operates only on the two-dimensional boundary mesh and returns sparse integer generators within a few seconds.

\subsection{Solver and Preconditioning}
\label{sec:solver}
\subsubsection{Solver}
Since the inductance operator is available only matrix-free (Section~\ref{sec:L-far}), every linear system must be solved iteratively. We therefore use SUNDIALS IDA, an implicit integrator for differential-algebraic equation (DAE) systems with an inexact Newton--Krylov core~\cite{knoll2004jfnk,hindmarsh2005sundials}. For the constant-resistivity benchmarks of this paper, the residual is linear in $\mathbf{T}$ and each Newton iteration converges in one step; the Newton framework is retained for planned nonlinear extensions such as a temperature- and field-dependent resistivity.

IDA~\cite{hindmarsh2005sundials,brown1994daspk} integrates~\eqref{eq:system} in residual form $\mathbf{F}\bigl(t,\mathbf{y},\dot{\mathbf{y}}\bigr)=\mathbf{0}$ with an adaptive backward differentiation formula (BDF), which adjusts order and step size after every accepted step to meet the prescribed tolerances. Each step requires the solution of a nonlinear system $G(\mathbf{y}_n)=\mathbf{0}$ by Newton iteration with the iteration matrix
\begin{equation}\label{eq:IDA-jacobian}
  \mathbf{J} \;=\; \frac{\partial \mathbf{F}}{\partial \mathbf{y}} + c_J\,\frac{\partial \mathbf{F}}{\partial \dot{\mathbf{y}}}
             \;=\; \mathbf{R} + c_J\,\mathbf{L},
\end{equation}
where the scalar $c_J \propto 1/\Delta t$ is supplied by IDA and changes whenever the step size or order changes~\cite{hindmarsh2005sundials}.

Each Newton step evaluates the residual $G(\mathbf{y}_n)=\mathbf{R}\,\mathbf{y}_n+\mathbf{L}\,\dot{\mathbf{y}}_n+d\mathbf{U}/dt$ once and solves $\mathbf{J}\,\Delta\mathbf{y}=-G$ with preconditioned GMRES (SUNDIALS \texttt{SPGMR}~\cite{sundials_gardner_2022}), which accesses the dense matrix $\mathbf{J}$ only through the matrix-free Jacobian--vector product (jtv) $\mathbf{J}\mathbf{v}=\mathbf{R}\mathbf{v}+c_J\,\mathbf{L}\mathbf{v}$. The preconditioner (see below) is applied from the left, so the GMRES stopping test acts on the preconditioned residual. IDA operates on the DOF vector, whereas $\mathbf{R}$, $\mathbf{L}$ and $d\mathbf{U}/dt$ act on the kept edges. The residual and the jtv therefore lift the DOF vector to the kept edges with $\mathbf{P}_\mathrm{cat}$ (Section~\ref{sec:bc-cohom}), apply the operators there, and project the result back with $\mathbf{P}_\mathrm{cat}^{\!\top}$. For scaling, the edge length is absorbed into the unknowns when $\mathbf{M}_{e\to f}$ is reduced to its $\pm 1$ incidence form, so every DOF has the unit ampere, and the diagonal scaling vectors of SPGMR are set to characteristic current and voltage magnitudes. This makes the tolerances physically interpretable and proved essential for robust convergence across mesh sizes~\cite{sundials_gardner_2022,knoll2004jfnk}.

Since each Newton step requires many GMRES iterations, the jtv dominates the runtime, and within it the inductance product~\eqref{eq:L-mvp} costs at least two orders of magnitude more than $\mathbf{R}\mathbf{v}$ or $d\mathbf{U}/dt$. How this cost splits between the MLFMM far-field pass and the sparse near+self product depends on the problem size. The MLFMM dominates for meshes of a few $10^3$ elements and the near+self product from about $10^5$ elements. In both cases, the performance is governed by the number of $\mathbf{L}\mathbf{v}$ products, i.e.\ by the number of Krylov iterations, which motivates the preconditioner described next.

\subsubsection{Preconditioner}
A preconditioner is an approximation $\mathbf{M}_\mathrm{prec}\approx\mathbf{J}^{-1}$ that is cheap to apply. With left preconditioning, GMRES solves $\mathbf{M}_\mathrm{prec}\mathbf{J}\,\Delta\mathbf{y}=-\mathbf{M}_\mathrm{prec}G$, which converges in far fewer iterations if $\mathbf{M}_\mathrm{prec}\mathbf{J}\approx\mathbf{I}$.

Of the sparse solvers and preconditioners benchmarked on our problems (CHOLMOD, UMFPACK, MUMPS, ParU, SuperLU, QDLDL, HYPRE AMS, and Jacobi), four backends are retained: CHOLMOD~\cite{cholmod} (sparse Cholesky), the fastest for symmetric positive definite (SPD) matrices; UMFPACK~\cite{umfpack} (sparse LU), the fastest fallback for non-SPD matrices; MUMPS~\cite{mumps2001} ($LDL^{\!\top}$ with null-pivot detection for the singular variants of Section~\ref{sec:gauging}), the fastest on multiple cores; and, for memory-limited cases, the factorization-free HYPRE AMS multigrid~\cite{kolev2009ams}, which exploits the edge-element structure at linear memory cost. No dense inverse is formed. The factorizing backends store sparse triangular factors with fill-reducing orderings and apply $\mathbf{M}_\mathrm{prec}\mathbf{v}$ by forward and backward substitution.

In the gauged system, $\mathbf{R}$ and the full $\mathbf{L}$ are SPD, so a preconditioner built from $\mathbf{R}$ and $c_J\mathbf{L}$ could be factored with CHOLMOD. The near+self block $\mathbf{L}_f^{\mathrm{n+s}}$, however, is not SPD, because the far-field part of $\mathbf{L}$ is omitted. At small time steps, $c_J$ grows, $c_J\mathbf{L}_f^{\mathrm{n+s}}$ dominates $\mathbf{R}$, and their sum becomes indefinite. It remains non-singular in practice, as its positive self-inductance diagonal keeps it full-rank, and can therefore be factored with UMFPACK.\footnote{The ungauged system (Section~\ref{sec:gauging}) and the lazy cohomology (Section~\ref{sec:bc-cohom}) make these matrices rank-deficient; they are then factored with the corrections of Section~\ref{sec:gauging}.} We derived four preconditioner variants that augment $\mathbf{R}$ with increasingly complete approximations of $c_J\mathbf{L}$:
\begin{itemize}[\IEEEsetlabelwidth{$\bullet$}\setlength{\IEEElabelindent}{0pt}\setlength{\itemsep}{0pt}\setlength{\topsep}{0pt}\setlength{\parsep}{0pt}\setlength{\partopsep}{0pt}]
\item \texttt{R\_ONLY} omits the inductive term, $\mathbf{M}_\mathrm{prec}\approx\mathbf{R}^{-1}$;
\item \texttt{R\_CJ\_LDIAG} adds the edge self-inductances, $\mathbf{M}_\mathrm{prec}\approx(\mathbf{R}+c_J\,\mathrm{diag}(\mathbf{L}))^{-1}$;
\item \texttt{R\_CJ\_LFDIAG} adds the face self-inductances, $\mathbf{M}_\mathrm{prec}\approx(\mathbf{R}+c_J\,\mathbf{M}_{e\to f}^{\!\top}\,\mathrm{diag}(\mathbf{L}_f^{\mathrm{n+s}})\,\mathbf{M}_{e\to f})^{-1}$;
\item \texttt{R\_CJ\_LFULL} includes the full near+self block, $\mathbf{M}_\mathrm{prec}\approx(\mathbf{R}+c_J\,\mathbf{M}_{e\to f}^{\!\top}\,\mathbf{L}_f^{\mathrm{n+s}}\,\mathbf{M}_{e\to f})^{-1}$.
\end{itemize}

These matrices are defined on the kept edges, whereas IDA requires the preconditioner in DOF space. For any of the four edge-space approximations $\tilde{\mathbf{J}}$ of $\mathbf{J}$, the change of basis $\mathbf{P}_\mathrm{cat}$ of Section~\ref{sec:bc-cohom} gives the DOF-space matrix by the congruence~\cite{bettini2017vinco}
\begin{equation}\label{eq:Mprec-DOF}
\mathbf{P}_\mathrm{cat}^{\!\top}\,\tilde{\mathbf{J}}\,\mathbf{P}_\mathrm{cat}
= \begin{bmatrix}
\mathbf{P}^{\!\top}\tilde{\mathbf{J}}\mathbf{P} & \mathbf{P}^{\!\top}\tilde{\mathbf{J}}\mathbf{H}\\[2pt]
\mathbf{H}^{\!\top}\tilde{\mathbf{J}}\mathbf{P} & \mathbf{H}^{\!\top}\tilde{\mathbf{J}}\mathbf{H}
\end{bmatrix},
\end{equation}
which is factored to obtain $\mathbf{M}_\mathrm{prec}\approx(\mathbf{P}_\mathrm{cat}^{\!\top}\tilde{\mathbf{J}}\mathbf{P}_\mathrm{cat})^{-1}$. Its diagonal blocks act on the interior and on the $N_\mathcal{H}$ cohomology DOFs, and its off-diagonal blocks couple them.

The first three variants are SPD and factored with CHOLMOD, and \texttt{R\_CJ\_LFULL} with UMFPACK. \texttt{R\_ONLY} does not depend on $c_J$ and is factored only once, whereas the other variants are refactored whenever IDA changes the step size or order. \texttt{R\_ONLY} is the default, since it is the only variant with $\mathbf{M}_\mathrm{prec}^{-1}\preceq\mathbf{J}$, which turns the left-preconditioned stopping test into a rigorous error bound.\footnote{With $\mathbf{M}_\mathrm{prec}^{-1}=\mathbf{R}$ and $\mathbf{J}-\mathbf{R}=c_J\,\mathbf{L}$ positive semi-definite, the error of the computed correction, $\mathbf{e}=\mathbf{J}^{-1}\mathbf{r}$ with $\mathbf{r}$ the true residual, is bounded by the monitored preconditioned residual: $\|\mathbf{e}\|_{\mathbf{R}}\le\|\mathbf{R}^{-1}\mathbf{r}\|_{\mathbf{R}}$ for every $c_J$. A small monitored residual therefore provably implies a small correction error at every step size. The $\mathbf{L}$-augmented variants use only part of $\mathbf{L}$, can overestimate $\mathbf{J}$ on some modes, and therefore carry no such one-sided bound.} It was also the fastest for the TEAM-7 simulations of Section~\ref{sec:results} and is used throughout. The four variants form an additive sequence $\tilde{\mathbf{J}} = \mathbf{R} + c_J\,\tilde{\mathbf{L}}$, with $\tilde{\mathbf{L}}$ ranging from $\mathbf{0}$ to the full near+self block. For such sequences, the deviation of the preconditioned operator from the identity is bounded by the \emph{omitted} part scaled by $c_J$~\cite{brown1994daspk,knoll2004jfnk}, so \texttt{R\_ONLY} performs best at large time steps, while the richer variants pay off only at small steps and large $c_J$, e.g.\ in high-frequency problems. The fastest variant therefore depends on the excitation time scales, the mesh size, and the available memory. RAT-EC provides all four as runtime options.

\section{Verification}\label{sec:verification}

This section verifies RAT-EC against closed-form solutions for two geometries in a homogeneous transient field: a conducting sphere, which tests the accuracy of the discretization on a solid body, and a conducting ring, which tests the treatment of multiply connected domains. Both references are analytical and thus free of experimental uncertainty.

The verification runs are configured for accuracy rather than speed. Conservative tolerances lead to 93--258 time steps per simulation (see the tables), whereas relaxed tolerances would give very similar results much faster. All simulations of this and the next section ran in Docker jobs on identical nodes of CERN's HTCondor batch system~\cite{htcondor_thain_2005} (AMD EPYC 9654 with up to 64 cores per job, 1.5\,TB of RAM, and one NVIDIA H200 GPU with 141\,GB), so all timings are directly comparable.

\subsection{Conducting Sphere}\label{sec:sphere}
\begin{figure}[!tbp]
    \centering
    \includegraphics[width=0.99\columnwidth]{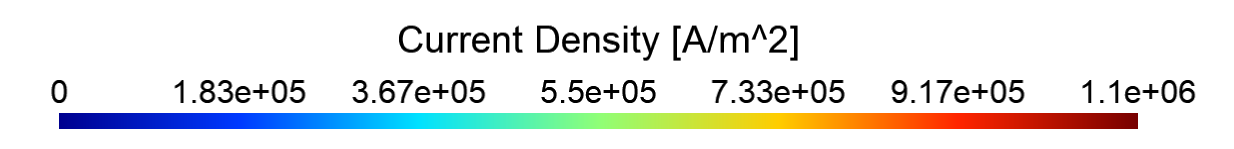}
    \par\nointerlineskip
    \begin{tikzpicture}
      \node[anchor=south west,inner sep=0] (img) at (0,0) {\includegraphics[width=0.7\columnwidth]{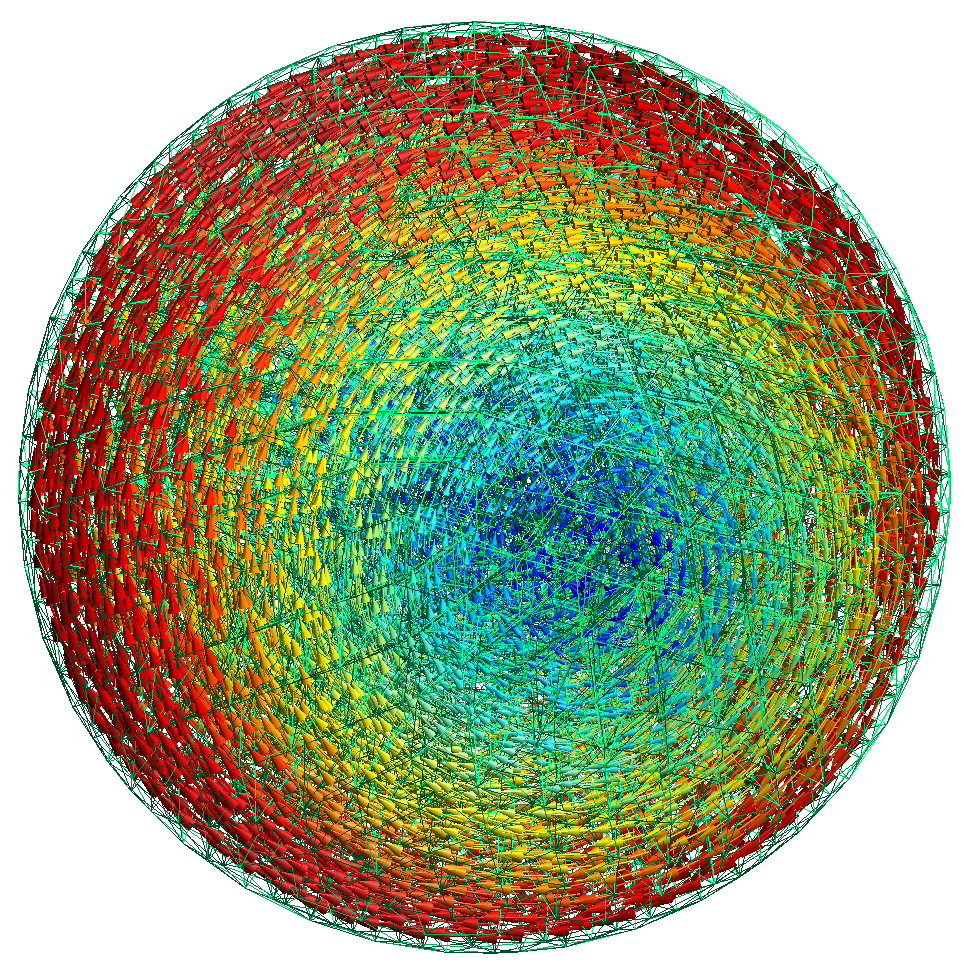}};
      \begin{scope}[overlay, shift={($(img.south east)+(1.0mm,4.0mm)$)}, x=8mm, y=8mm, line width=0.5pt,
                    -{Stealth[length=1.5mm,width=1.0mm]}, font=\footnotesize]
        \draw (0,0) -- (0.0117,0.9863)   node[above,inner sep=1pt]         {$y$};
        \draw (0,0) -- (-0.9607,-0.0345) node[left,inner sep=1pt]          {$z$};
        \draw (0,0) -- (0.2773,-0.1614)  node[below right,inner sep=0.5pt] {$x$};
      \end{scope}
    \end{tikzpicture}
    \caption{Eddy-current density $|\mathbf{j}|$ [A/m$^2$] in the conducting sphere under a step in $\dot B_x$, computed with RAT-EC on mesh S3 at $t=40$\,ms ($\approx 14\,\tau_1$, steady-state plateau); arrows at the tetrahedron centroids, mesh overlaid, visualized with Gmsh~\cite{Gmsh_ijnme}. The current forms the toroidal vortex predicted by theory, without any mesh of the surrounding air.}
    \label{fig:sphere_eddy}
\end{figure}

A solid copper sphere of radius $a = 20$\,mm and conductivity $\sigma = 1/\rho_{\mathrm{Cu}} = 5.65\times 10^{7}$\,S/m is exposed at $t=0$ to a uniform field with a step in $\dot B_x$ to $2$\,T/s. The sphere is meshed with the RAT-distmesh module~\cite{rat_group}, based on the distance-function mesher of~\cite{persson2004distmesh}, from the signed distance $|\mathbf{r}|-a$ and a single target edge length $h$. Four resolutions S1--S4 (Table~\ref{tab:sphere_verification}) span nearly three orders of magnitude in DOF count. Each hexahedron of the underlying mesh is split into 24 tetrahedra by a symmetric face-centered subdivision that ensures conforming interior faces regardless of the diagonal orientation. Fig.~\ref{fig:sphere_eddy} shows the eddy currents on mesh S3, viewed along the field axis. They form the expected toroidal vortex, symmetric within the discretization error about the sphere diameter aligned with $\dot B_x$. Relative and absolute tolerances of $10^{-4}$ and $10^{-6}$ ($10^{-5}$ on S4), without a limit on the step size, leave the step selection to the error control, which resolves the slowest diffusion mode with 107--258 steps.

Following the classical treatment of magnetic diffusion into a conducting sphere~\cite{smythe1968static,landau1984electrodynamics}, separation of variables in spherical coordinates yields a closed-form Bessel-series expansion of the eddy-current density. Integrating $\rho\,|\mathbf{j}|^2$ over the sphere gives the total ohmic power
\begin{equation}\label{eq:sphere_p}
P(t) = \frac{12}{\pi^3}\,\sigma\,a^5\,\dot B^2 \sum_{n\geq 1} \frac{\bigl(1-e^{-t/\tau_n}\bigr)^2}{n^4},
\quad
\tau_n = \frac{\mu_0 \sigma a^2}{n^2\pi^2},
\end{equation}
with the steady-state plateau
\begin{equation}\label{eq:sphere_pinf}
P_{\infty} = \frac{2\pi}{15}\,\sigma\,a^5\,\dot B^2.
\end{equation}
The slowest mode, $\tau_1 = \mu_0\sigma a^2/\pi^2$, sets the diffusion time scale. At $t_\mathrm{end} \approx 14\,\tau_1$ ($40$\,ms), it has decayed by $e^{-14}\sim 10^{-6}$, so $P_\infty$ is reached within $0.01\,\%$.

\begin{figure}[!tbp]
    \centerline{\includegraphics[width=\columnwidth]{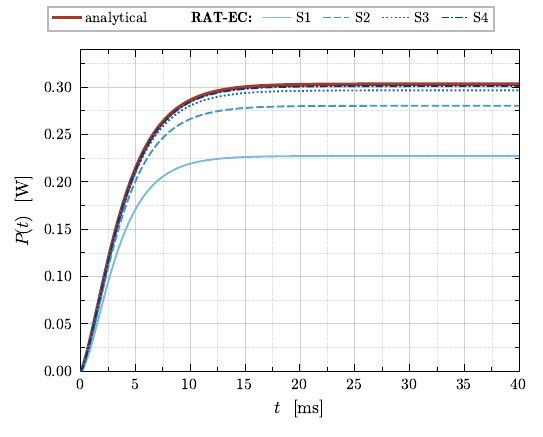}}
    \vspace{-7pt}
    \caption{Dissipated power $P(t)$ in the conducting sphere under a step in $\dot B_x$. Analytical reference \eqref{eq:sphere_p} (solid red), approaching the plateau \eqref{eq:sphere_pinf}, and RAT-EC solutions on meshes S1--S4 (coarse to fine with errors in Table~\ref{tab:sphere_verification}). The computed power converges monotonically to the analytical solution, and the intermediate meshes already capture both the rise and the plateau.}
    \label{fig:sphere_convergence}
\end{figure}

Fig.~\ref{fig:sphere_convergence} compares the RAT-EC solutions on S1--S4 with the analytical $P(t)$ of \eqref{eq:sphere_p}. The solutions converge monotonically, with the plateau error decreasing from $\approx 25\,\%$ on S1 to $\approx 0.6\,\%$ on S4. The diffusion time constant $\tau_1$, obtained from an exponential fit of each simulated $P(t)$, converges likewise, with an error $\epsilon_\tau$ decreasing from $\approx 6.5\,\%$ to below $0.01\,\%$ (Table~\ref{tab:sphere_verification}).

\begin{table*}[!tbp]
    \caption{Sphere Verification: Mesh Sizes, DOF Counts After Gauging, Plateau Error $\epsilon_{P_\infty}$ and Time-Constant Error $\epsilon_\tau$ Relative to closed form, and Wall-Clock Times of the One-Time Assembly $t_{\mathrm{asm}}$ (Mesh Setup and $\mathbf{L}_{\mathrm{n+s}}$) and the Time Integration $t_{\mathrm{solve}}$ on the Node of Section~\ref{sec:verification}}\label{tab:sphere_verification}
    \centering
    \begin{tabular}{cccccccccc}
    \hline
    Mesh & $h$ [mm] & \#tets & DOF & $\epsilon_{P_\infty}$ [\%] & $\epsilon_\tau$ [\%] & \#steps & $t_{\mathrm{asm}}$ [s] & $t_{\mathrm{solve}}$ [s] & $t_{\mathrm{solve}}/$step [s] \\
    \hline
    S1 & 10  & 160     & 121     & 25.0 & 6.5   & 107 & 0.5 & 21   & 0.20 \\
    S2 &  5  & 1\,280  & 1\,121  & 7.5  & 1.1   & 160 & 0.6 & 39   & 0.24 \\
    S3 &  2  & 10\,240 & 9\,601  & 2.1  & 0.11  & 231 & 4.4 & 98   & 0.42 \\
    S4 &  1  & 81\,920 & 79\,361 & 0.6  & 0.007 & 258 & 264 & 582  & 2.26 \\
    \hline
    \end{tabular}
    \end{table*}

\subsection{Conducting Ring}\label{sec:ring}

\begin{figure}[!tbp]
    \centering
    \includegraphics[width=0.99\columnwidth]{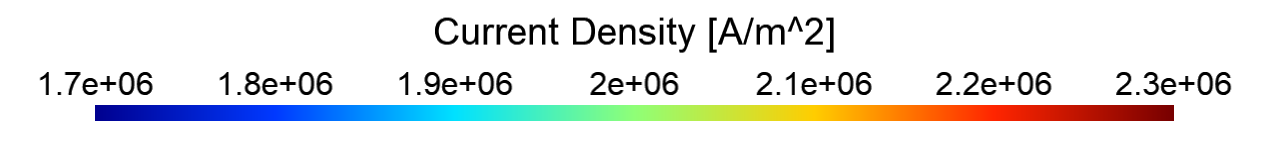}
    \par\nointerlineskip
    \includegraphics[width=0.7\columnwidth]{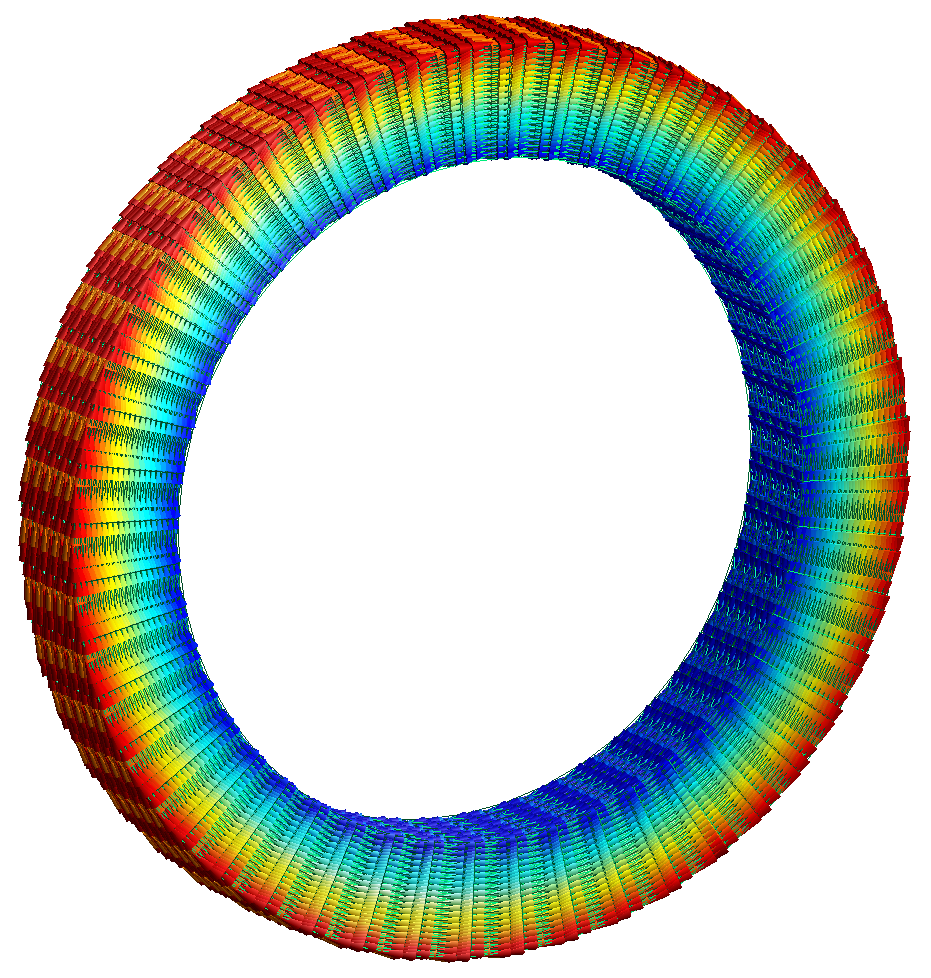}
    \caption{Eddy-current density $|\mathbf{j}|$ [A/m$^2$] in the conducting ring under a step in $\dot B_x$, computed with RAT-EC on the finest mesh R4 at $t=25$\,ms ($\approx 10\,\tau$, steady-state plateau); arrows at the tetrahedron centroids, mesh overlaid, visualized with Gmsh~\cite{Gmsh_ijnme}. The current circulates around the hole as a single loop, which confirms the topology-aware boundary condition.}
    \label{fig:ring_eddy}
\end{figure}

A copper ring with the conductivity of the sphere, an inner radius $r_i = 30$\,mm, an outer radius $r_o = 40$\,mm and a height $h = 10$\,mm is exposed to a uniform field with a step in $\dot B_x$ to $2$\,T/s along its axis. The ring is meshed by extruding a two-dimensional (2D) annular RAT-distmesh~\cite{rat_group} triangulation over the height $h$ in uniform hexahedral layers, each split into 24 tetrahedra as for the sphere. Four resolutions R1--R4 (Table~\ref{tab:ring_verification}) span nearly three orders of magnitude in DOF count. Fig.~\ref{fig:ring_eddy} shows the eddy currents on mesh R4. They form a single loop around the hole, a topological effect that the simply connected sphere cannot test and that RAT-EC captures with one cohomology DOF per hole (Section~\ref{sec:bc-cohom})~\cite{pellikka2013cohomology,dlotko2014lazy}.

In the thin-ring approximation with uniform current over the cross-section, the resistance, self-inductance~\cite{dengler2016selfinductance,grover1946inductance}, effective flux-collection area and time constant are
\begin{equation}\label{eq:ring-lumped}
\begin{aligned}
R &= \rho\,\frac{2\pi a_m}{A_{\mathrm{cs}}}, &
L &\approx \mu_0\,a_m\left[\ln(8 a_m/r_{\mathrm{eq}}) - 1.75\right], \\
A_{\mathrm{eff}} &= \frac{\pi (r_o^3 - r_i^3)}{3(r_o - r_i)}, &
\tau &= L/R,
\end{aligned}
\end{equation}
where $a_m = (r_i+r_o)/2$, $A_{\mathrm{cs}} = (r_o-r_i)\,h$, and $r_{\mathrm{eq}} = \sqrt{A_{\mathrm{cs}}/\pi}$ is the radius of the equivalent circular cross-section. These parameters form a series $R$--$L$ circuit driven by the EMF $A_{\mathrm{eff}}\,\dot B$, with the dissipated power
\begin{equation}\label{eq:ring_p}
P_{\mathrm{ana}}(t) = P_{\mathrm{ss}}\,\bigl(1-e^{-t/\tau}\bigr)^2,
\quad
P_{\mathrm{ss}} = (A_{\mathrm{eff}}\,\dot B)^2 / R.
\end{equation}
This model is approximate, as it neglects the eddy currents within the cross-section and the distributed self-inductance of the thick ring. The comparison therefore verifies the treatment of the topology and the $L/R$ response rather than the absolute accuracy.

\begin{figure}[!tbp]
    \centerline{\includegraphics[width=\columnwidth]{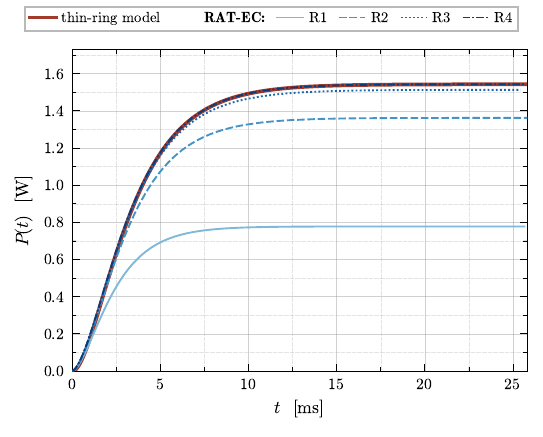}}
    \vspace{-5pt}
    \caption{Dissipated power $P(t)$ in the conducting ring under a step in $\dot B_x$: thin-ring reference \eqref{eq:ring_p} (solid red) and RAT-EC solutions on meshes R1--R4 (coarse to fine with errors in Table~\ref{tab:ring_verification}). Refinement recovers both the plateau and the $L/R$ rise time, which verifies the topology-aware constraints on a domain with a hole. Since the reference is approximate, this case tests topology and time constant rather than absolute accuracy.}
    \label{fig:ring_convergence}
\end{figure}

Fig.~\ref{fig:ring_convergence} compares the RAT-EC solutions on R1--R4 with \eqref{eq:ring_p}. The solutions converge monotonically, with the plateau error decreasing from $\approx 50\,\%$ on R1 to $\approx 0.2\,\%$ on R4 and the error $\epsilon_\tau$ of the fitted time constant from $\approx 30\,\%$ to $\approx 2.3\,\%$ (Table~\ref{tab:ring_verification}). This difference is consistent with the reference. The plateau depends only on the resistance and the effective area, which the lumped model captures accurately, whereas $\tau$ inherits the error of the thin-ring inductance formula in~\eqref{eq:ring-lumped}. The relative/absolute tolerances are $10^{-4}$/$10^{-6}$ for R1--R2, $10^{-4}$/$10^{-5}$ for R3 and $10^{-3}$/$10^{-4}$ for R4, relaxed on the finer meshes to limit the number of steps.

\begin{table*}[!tbp]
    \caption{Ring Verification: Mesh Sizes, DOF Counts After Gauging and Boundary Conditions, Plateau Error $\epsilon_{P_{\mathrm{ss}}}$ and Time-Constant Error $\epsilon_\tau$ Relative to the Thin-Ring Reference, and Wall-Clock Times of the One-Time Assembly $t_{\mathrm{asm}}$ and the Time Integration $t_{\mathrm{solve}}$ on the Node of Section~\ref{sec:verification}}\label{tab:ring_verification}
    \centering
    \begin{tabular}{ccccccccc}
    \hline
    Mesh & \#tets & DOF & $\epsilon_{P_{\mathrm{ss}}}$ [\%] & $\epsilon_\tau$ [\%] & \#steps & $t_{\mathrm{asm}}$ [s] & $t_{\mathrm{solve}}$ [s] & $t_{\mathrm{solve}}/$step [s] \\
    \hline
    R1 & 240      & 181      & 49.6 & 29.6 & 126 & 0.6 & 22  & 0.18 \\
    R2 & 1\,320   & 1\,057   & 11.9 & 7.5  & 137 & 0.6 & 27  & 0.20 \\
    R3 & 19\,008  & 17\,665  & 2.2  & 3.2  & 137 & 27  & 52  & 0.38 \\
    R4 & 109\,440 & 104\,833 & 0.16 & 2.3  & 93  & 386 & 149 & 1.60 \\
    \hline
    \end{tabular}
    \end{table*}

\section{Validation}\label{sec:results}

This section validates RAT-EC on TEAM-7, the standard measurement benchmark for 3D eddy currents, at both published frequencies, $f = 50$\,Hz and $f = 200$\,Hz, and compares it with an independent finite-element solution.

\begin{figure}[!tbp]
    \centerline{\includegraphics[width=\columnwidth]{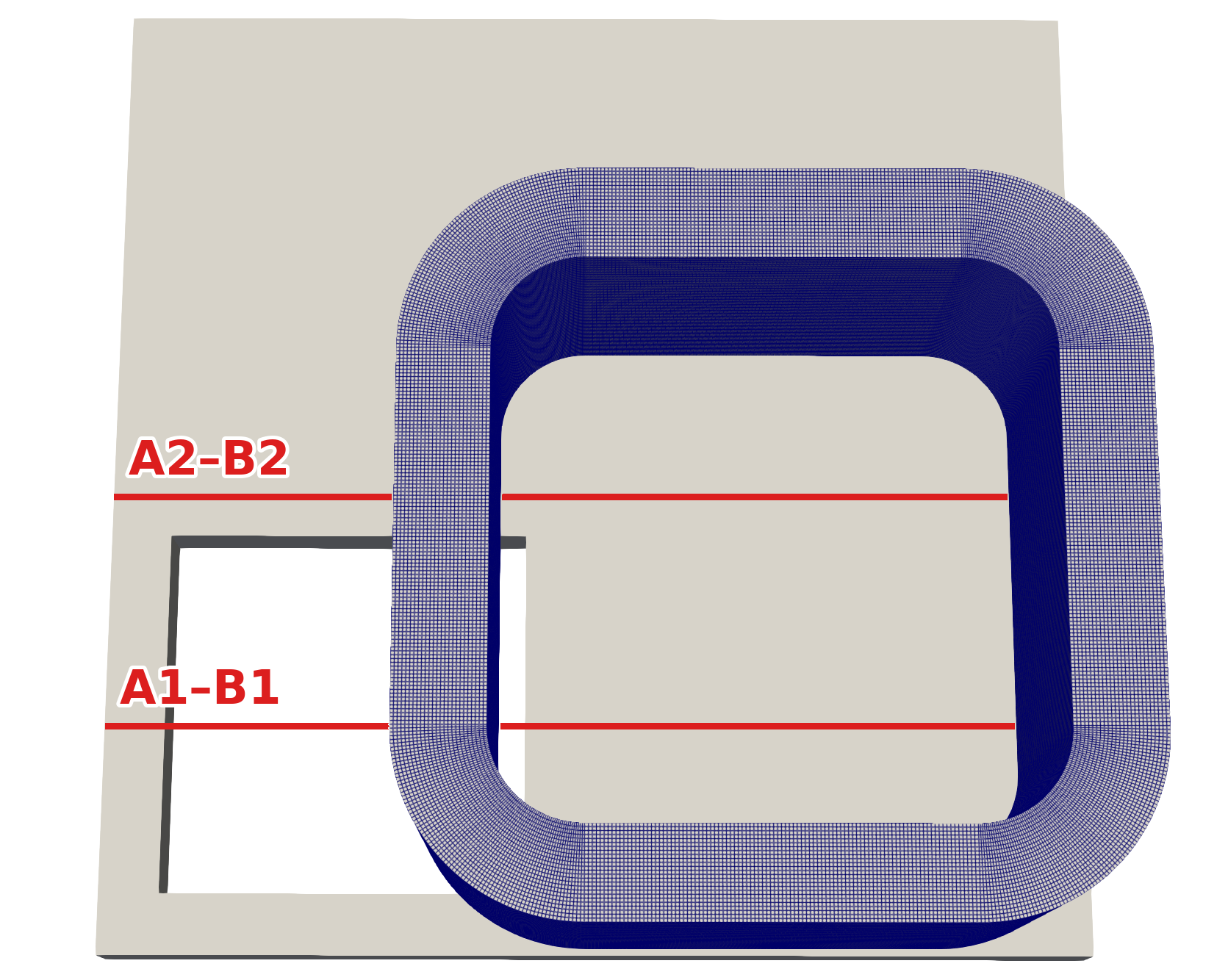}}
    \caption{TEAM-7 setup: aluminium plate with an off-centre rectangular hole below a rectangular coil with sinusoidal current. The red lines mark the $B_z$ measurement lines A1--B1 and A2--B2 in the air between plate and coil (visualized with ParaView~\cite{paraview}). The off-centre hole removes all symmetry planes, so the full 3D current distribution must be resolved. Only the plate is discretized. The coil enters as a prescribed line current. No air meshed.}
    \label{fig:team7_setup}
\end{figure}

TEAM-7~\cite{fujiwara1990team7} consists of an aluminium plate ($\sigma = 3.526\times 10^{7}$\,S/m) with an off-centre rectangular hole, excited by a rectangular coil with a sinusoidal current (Fig.~\ref{fig:team7_setup}). Mesh convergence and runtime are discussed for 50\,Hz. The two $B_z$ measurement lines of the benchmark, A1--B1 and A2--B2 (red in Fig.~\ref{fig:team7_setup}), run in the air at $z = 34$\,mm, midway between the plate top surface and the coil bottom.

\subsection{Initial Conditions and Time-Stepping}\label{sec:setup_ic}
RAT-EC works in the time domain since the magnet transients it targets, such as quenches and ramps, do not permit a frequency-domain treatment, although TEAM-7's sinusoidal regime would. The periodic steady state of the benchmark is therefore reached by transient integration. Since IDA's consistent initial-condition (IC) solver fails if the sinusoid is applied abruptly at $t=0$, the coil current is ramped up as
\begin{equation}\label{eq:ramp}
I(t) = I_0\cos(\omega t)\,\bigl(1-\cos(\pi t/t_\mathrm{ramp})\bigr)/2,
\end{equation}
with the angular frequency $\omega = 2\pi f$ and $t_\mathrm{ramp}=1/f$, starting from the consistent state $I=0$, $\dot I=0$. The simulation runs until the dissipated power $P(t)$ in the plate is periodic, i.e.\ until successive maxima differ by less than $0.1\,\%$, which takes five periods at $50$\,Hz and about eight at $200$\,Hz. More periods are needed at the higher frequency because the eddy-current transient is governed by the $L/R$ time constant of the conductor, which does not decrease as the source frequency increases. Only the final period is used for the comparison. The preceding swing-in transient replaces an explicit consistent-IC computation.

The step size is limited to $1/(100\,f)$, i.e.\ at least 100 steps per period, and the relative and absolute tolerances are $10^{-5}$ and $10^{-5}$\,A\footnote{After edge-length scaling (Section~\ref{sec:solver}), the unknowns are face-current contributions rather than electric vector potentials. Hence, absolute tolerances are set in amperes.}, respectively. The absolute tolerance is loose, but a tighter value would dominate during the low-amplitude ramp and enforce much smaller steps without improving the steady-state accuracy. The initial steps are about $1/(1000\,f)$, small enough for the ramp without noticeably increasing the runtime.

\subsection{Meshing}\label{sec:setup_meshing}
The coil (Fig.~\ref{fig:team7_setup}) is meshed parametrically with structured hexahedra by the RAT-models module. The racetrack centerline, with straight legs and circular-arc corners, is sampled at the axial element size $h$, and the rectangular cross-section is subdivided perpendicular to the path.

The plate is meshed with RAT-distmesh~\cite{rat_group}, a C++ port of the distance-function mesher of Persson and Strang~\cite{persson2004distmesh}. A 2D Delaunay triangulation~\cite{delaunay1934} of the plate footprint, generated from the signed distance function and a target element-size field, is extruded through the 19\,mm thickness in uniform hexahedral layers, and each hexahedron is split into 24 tetrahedra as in Section~\ref{sec:sphere}.

\begin{figure}[!t]
    \centerline{\begin{overpic}[width=0.9\columnwidth]{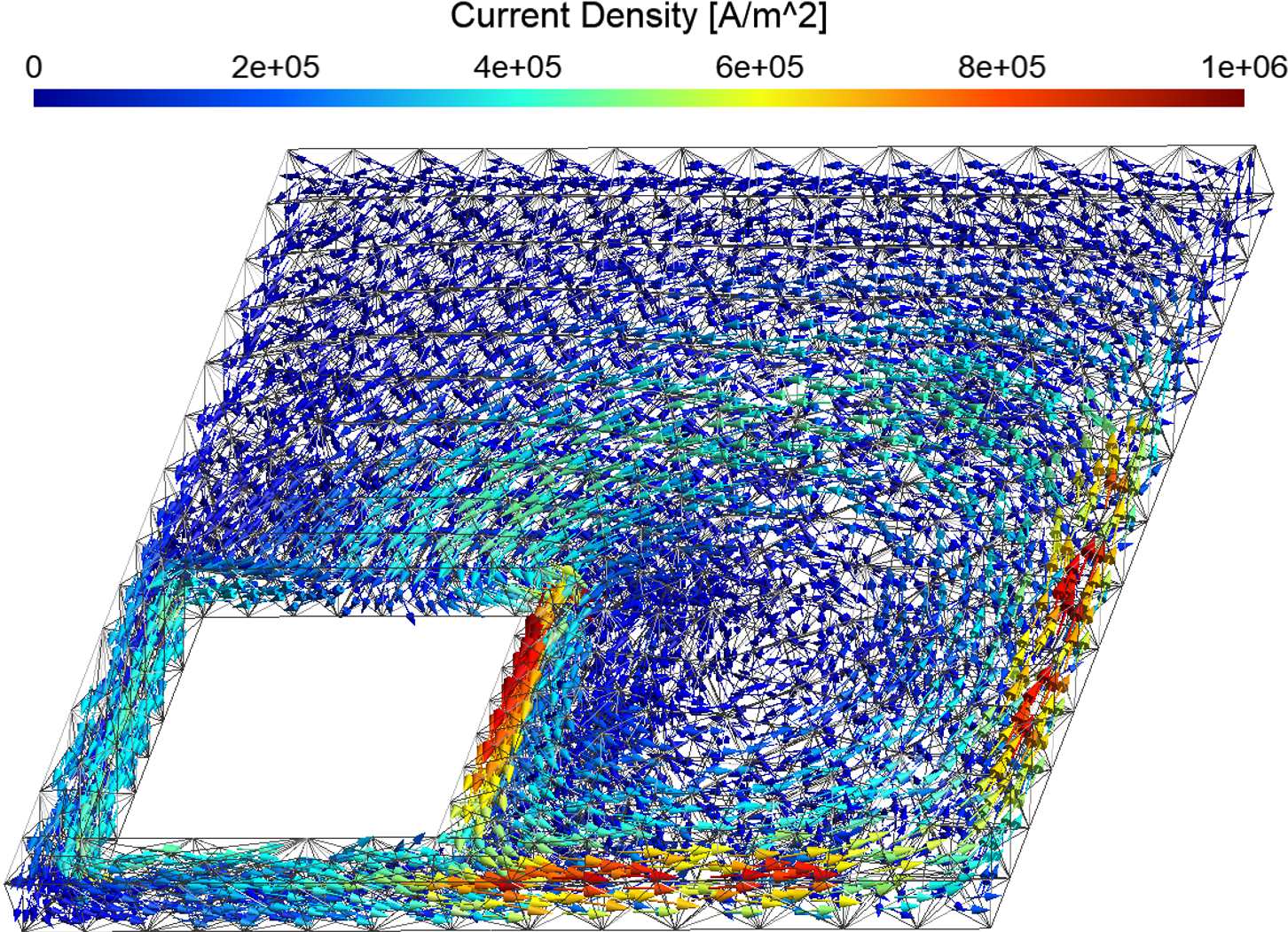}\panel{1}{55}{a}\end{overpic}}
    \vspace{10pt}
    \centerline{\begin{overpic}[width=0.9\columnwidth]{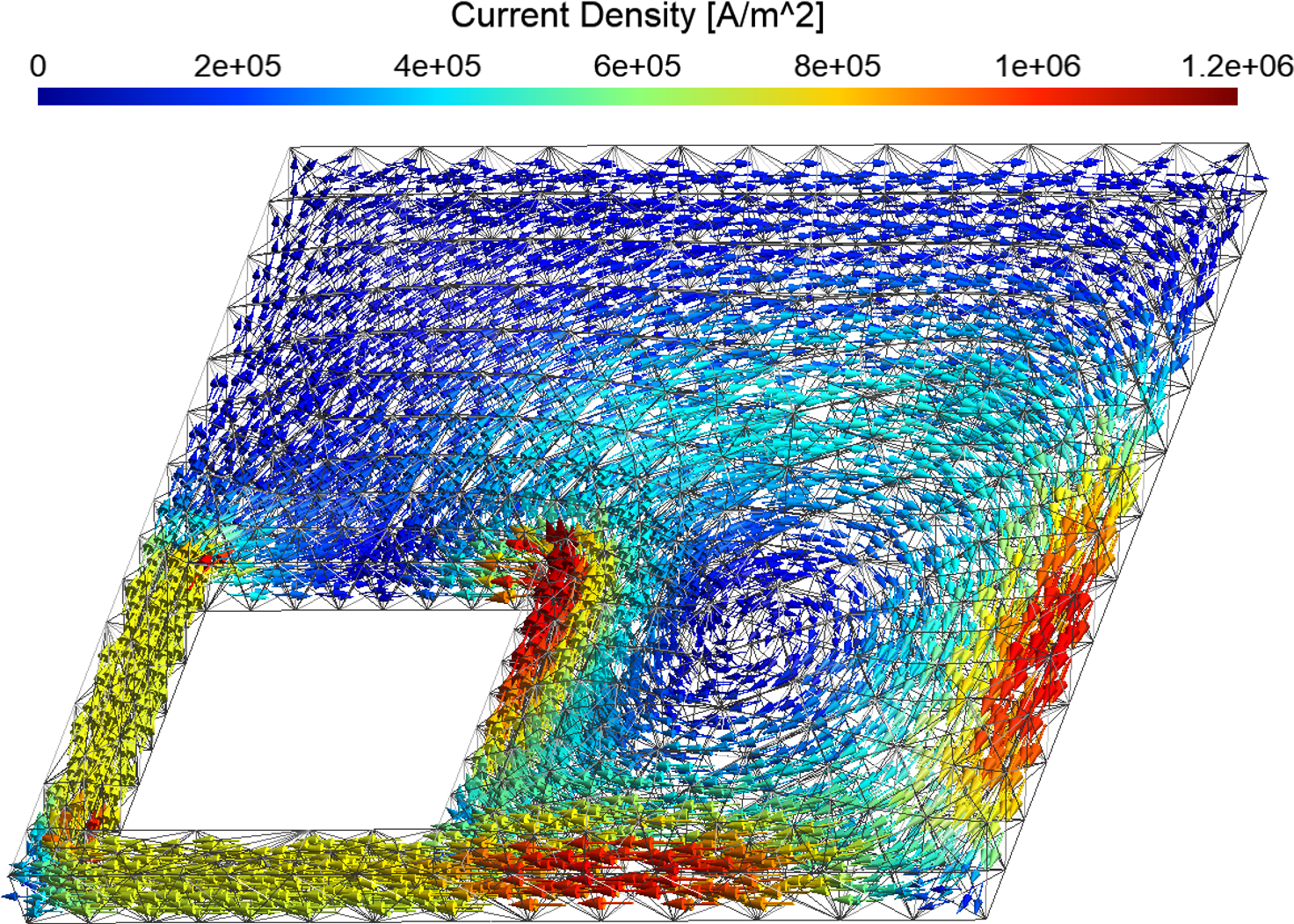}\panel{1}{54}{b}\end{overpic}}
    \caption{Current density $|\mathbf{j}|$ [A/m$^2$] in the TEAM-7 plate at 50\,Hz, computed with RAT-EC on the coarsest mesh M1: (a) $\omega t = 0^{\circ}$, coil current at peak; (b) $\omega t = 90^{\circ}$, coil current at zero crossing. Arrows at the tetrahedron centroids, mesh overlaid, visualized with Gmsh~\cite{Gmsh_ijnme}. In (a), the current concentrates at the face toward the coil; in (b), the free eddy-current response loops around the hole. Both patterns are resolved already on this coarse mesh.}
    \label{fig:team7_eddy}
\end{figure}

\begin{figure*}[!t]
    \noindent\begin{overpic}[width=\columnwidth]{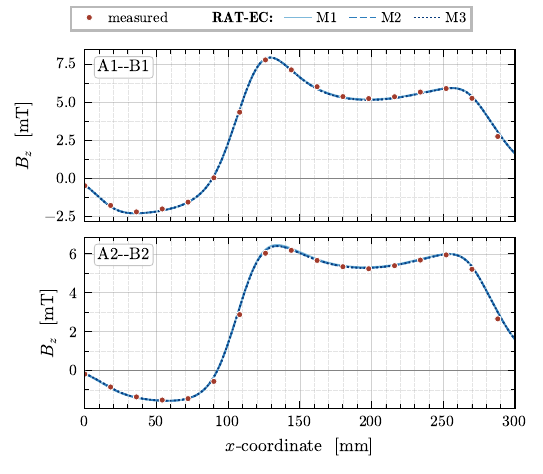}\panel{1}{81}{a}\end{overpic}\hfill
    \begin{overpic}[width=\columnwidth]{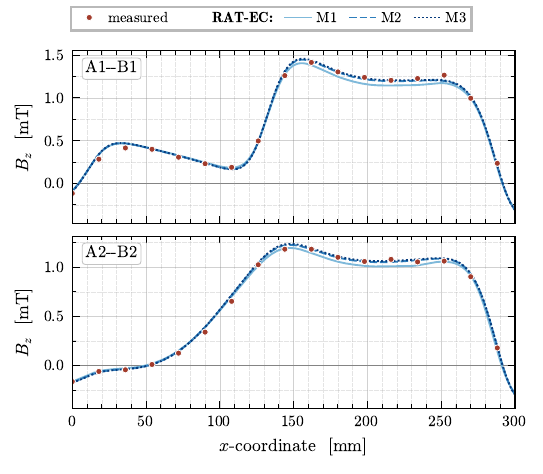}\panel{1}{81}{b}\end{overpic}\par
    \vspace{-7pt}
    \caption{Magnetic flux density $B_z$ along the TEAM-7 measurement lines A1--B1 (top) and A2--B2 (bottom) at 50\,Hz: RAT-EC on meshes M1--M3 (coarse to fine) and the Fujiwara--Nakata reference (red markers). (a)~$\omega t = 0^{\circ}$, coil current at peak (left); (b)~$\omega t = 90^{\circ}$, coil current at zero crossing (right). In (a), the coil dominates the field, so all meshes coincide and follow the measurement, this validates the current line model. In (b), the field is that of the eddy currents alone; the solutions converge under refinement, and the remaining offset matches that of a converged commercial FE model (deviations in Table~\ref{tab:team7_err}).}
    \label{fig:b_50Hz}
\end{figure*}

\subsection{Mesh Sizing and Convergence}\label{sec:setup_meshsizing}
A convergence study over $h\in[0.5,25]$\,mm shows that the peak $|B_z|$ along the measurement lines deviates by less than $5\times 10^{-5}$ from its converged value at $h=2$\,mm. This resolution ($234\,000$ segments) is used for the coil. The CLM is evaluated three times: (i) at $170$ points along each measurement line~\cite{fujiwara1990team7}, which gives the coil contribution to $B_z$ as a smooth curve at a finer spacing than the published Fujiwara--Nakata points; (ii) at all Gauss points of the plate mesh to precompute the right-hand side $\mathbf{U}$, which is scaled with the coil current at every time step; and (iii) at the same line points after the simulation, for the eddy-current solution, to obtain the eddy contribution to $B_z$, which is added to (i) to give the total field. At $h=2$\,mm, the line evaluation (i) takes $\sim$1.5\,s ($\sim$132\,s at $h=0.5$\,mm), and the Gauss-point evaluation (ii) takes less than $0.2$\,s on the GPU even on the finest plate mesh (Table~\ref{tab:mesh_setup}).

\begin{table}[!tbp]
\caption{TEAM-7 Mesh Sizes and Inductance-Block Setup Costs}\label{tab:mesh_setup}
\centering
\setlength{\tabcolsep}{4pt}
\begin{tabular}{cccccc}
\hline
\multirow{2}{*}{} & \multirow{2}{*}{DOF$^{*}$} & \multicolumn{3}{c}{$\mathbf{L}_{\mathrm{n+s}}$ matrix} & \multirow{2}{*}{CLM$^{\dagger}$} \\
\cline{3-5}
Mesh & & assembly$^{\dagger}$ & size & $\langle\#\mathrm{near}\rangle$ & \\
\hline
M1 & 4{,}901    & 0.3\,s &  22\,MB & 220 & 0.13\,s \\
M2 & 51{,}989   &  10\,s & 668\,MB & 742 & 0.13\,s \\
M3 & 175{,}249  &  14\,s & 900\,MB & 298 & 0.17\,s \\
\hline
\multicolumn{6}{l}{\footnotesize $^{*}$After tree--cotree gauge and boundary-condition enforcement.} \\
\multicolumn{6}{l}{\footnotesize $^{\dagger}$Wall-clock on the node of Section~\ref{sec:verification}.} \\
\end{tabular}
\end{table}

Three plate meshes M1--M3 with 5\,k to 175\,k DOF demonstrate convergence (Table~\ref{tab:mesh_setup}). All have the same in-plane grading. With a gradation factor of $0.3$~\cite{persson2004distmesh}, the target element size grows smoothly from a minimum at the hole edge, where the highest accuracy is needed, to a maximum at the outer boundary. The meshes differ in the number of layers and in the minimum and maximum in-plane sizes: one layer with $h\in[19,\,40]$\,mm for M1, two layers with $h\in[6,\,20]$\,mm for M2, and three layers with $h\in[5,\,12]$\,mm for M3. Table~\ref{tab:mesh_setup} lists the one-off setup costs: the assembly time and storage of $\mathbf{L}_\mathrm{n+s}$, the average number of near-field elements per source element $\langle\#\mathrm{near}\rangle$, and the CLM precomputation time. Mesh generation, gauging and cohomology computation take negligible time in comparison.

\subsection{COMSOL Reference}\label{sec:setup_comsol}
As an independent FE reference, TEAM-7 is also simulated with the official COMSOL Multiphysics~6.4 model of the COMSOL Application Gallery~\cite{comsol_team7}, using the frequency-domain formulation of the AC/DC module. The maximum element sizes of the plate and of the coil are refined jointly until the ohmic power in the plate changes by less than $1\,\%$ between refinement steps. At 50\,Hz, this is reached at about $6\times 10^5$ DOF with both sizes set to $5$\,mm, which serves as the COMSOL reference in Table~\ref{tab:team7_err}.

\subsection{Results}\label{sec:team7_results}

Fig.~\ref{fig:team7_eddy} shows the simulated eddy-current density in the plate at the phase angles $\omega t = 0^{\circ}$ and $90^{\circ}$. In both cases, the currents circulate around the off-centre hole in the asymmetric pattern characteristic of TEAM-7. At $\omega t = 0^{\circ}$ (panel~(a), coil current at peak), the currents concentrate at the top surface of the plate, a skin effect that is visible even on the coarse mesh M1 with its limited resolution through the thickness. At $\omega t = 90^{\circ}$ (panel~(b), coil current at zero crossing), the whole plate carries a vortex around the hole, the free eddy-current response without an instantaneous source contribution.

For a quantitative comparison, $B_z$ is sampled along A1--B1 and A2--B2 at the standard phase angles $\omega t = 0^{\circ}$ and $90^{\circ}$. At $\omega t = 0^{\circ}$ (Fig.~\ref{fig:b_50Hz}(a)), the coil field dominates $B_z$ in the measurement plane and the in-phase eddy-current contribution is small, so this phase mainly validates the CLM that supplies $\mathbf{a}_s$. As expected, the eddy-current discretization has little influence. The results on M1--M3 coincide and follow the reference, with an RMS deviation of $\approx$1.5\,\% on A1--B1 and $\approx$2.4\,\% on A2--B2 and a signed maximum deviation $\Delta B_z^{\max}$ of $\approx$3.8\,\% and $\approx$6.2\,\%, respectively.

The discriminating case is $\omega t = 90^{\circ}$ (Fig.~\ref{fig:b_50Hz}(b)), where the coil current vanishes and $B_z$ is determined by the eddy currents alone. This comparison is more demanding, since it probes the small eddy-current field directly and is sensitive to the mesh resolution and to noise in the reference data. RAT-EC converges under mesh refinement, and the remaining deviation from the Fujiwara--Nakata data is comparable to that of the converged commercial FE model (COMSOL AC/DC~\cite{comsol_team7}, Table~\ref{tab:team7_err}). RAT-EC reaches this accuracy already on mesh M2 with ${\sim}52$\,k DOF, an order of magnitude fewer unknowns than the FE reference, with only the plate meshed.

\begin{table}[!tbp]
\caption{$B_z$ Deviation From the TEAM-7 Reference at $\omega t = 90^{\circ}$; RMS and Signed $\Delta B_z^{\max}$ at the Worst Point (Positive = Over-Shoot)}\label{tab:team7_err}
\centering
\begin{tabular}{ccccc}
\hline
 & \multicolumn{2}{c}{RMS$^{*}$} & \multicolumn{2}{c}{$\Delta B_z^{\max\,*}$ (signed)} \\
\cline{2-5}
 Model & A1--B1 & A2--B2 & A1--B1 & A2--B2 \\
\hline
RAT-EC mesh M1 & 3.4 & 2.9 & $-6.6$ & $-6.0$ \\
RAT-EC mesh M2 & 2.2 & 2.2 & $-4.7$ & $+5.3$ \\
RAT-EC mesh M3 & 2.2 & 2.4 & $-4.3$ & $+5.5$ \\
COMSOL reference$^{\dagger}$ & 3.7 & 2.5 & $-7.5$ & $-5.0$ \\
\hline
\multicolumn{5}{l}{\footnotesize $^{*}$\% of the panel-peak $|B_z|$ of the measured reference.} \\
\multicolumn{5}{l}{\footnotesize $^{\dagger}$$\sim$6$\times$10$^{5}$ DOF FE model~\cite{comsol_team7}.} \\
\end{tabular}
\end{table}

Table~\ref{tab:perf} summarizes the wall-clock costs. In the periodic regime, a time step takes about $0.1$\,s on M1, $0.5$\,s on M2 and $1.8$\,s on M3. The swing-in transient dominates the total, e.g.\ $22$\,min compared with $3.6$\,min for the final period on M3. About one Newton iteration per step suffices, as the residual is linear at the constant resistivity prescribed by the benchmark~\cite{fujiwara1990team7}. More iterations occur only during the ramp, while the BDF order and step size adapt and the predictor lags behind the rapidly changing source. The number of Krylov iterations is low and depends only weakly on the mesh ($\langle$GMRES/step$\rangle = 19, 23, 25$ for M1--M3), owing to the preconditioner (Section~\ref{sec:solver}). The cost of each GMRES iteration is dominated by the inductance product, whose main part shifts from the MLFMM far field on M1 to the near+self block on M3. Both parts run on the GPU, which gives a speed-up of about one order of magnitude over the CPU-only version, depending on the mesh and consistent with Fig.~\ref{fig:fmm-scaling}, and makes M3 tractable on a single GPU node. All reported timings include GPU acceleration.

\begin{figure}[!tbp]
    \centerline{\includegraphics[width=\columnwidth]{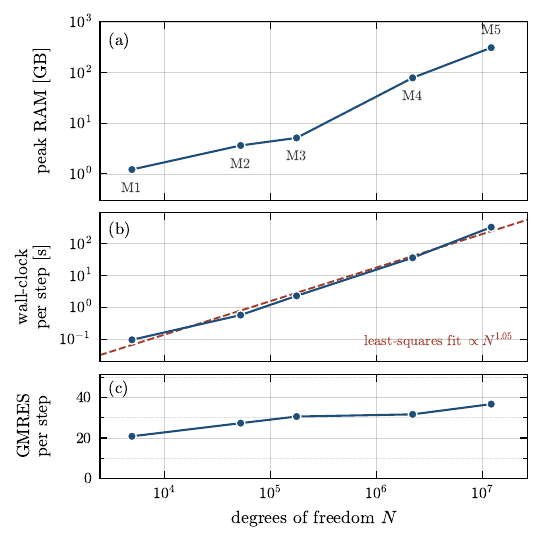}}
    \vspace{-6pt}
    \caption{Resource scaling of RAT-EC on the TEAM-7 plate meshes M1--M5 (4.9\,k to 12\,M DOF) on the hardware of Section~\ref{sec:verification}: (a)~peak host memory of the process (kernel high-water mark); (b)~wall-clock time and (c)~GMRES iterations per step, averaged over the ramp period, with a least-squares power-law fit in~(b). The memory follows the near-field storage set by the octree rather than the DOF count alone, the time per step grows slightly faster than linearly, and the Krylov count grows only weakly.}
    \label{fig:scaling}
\end{figure}

To test the limits, two finer plate meshes, M4 and M5 with 2.2 and 12.0 million DOF, respectively, were each integrated over the ramp period on the same hardware (Fig.~\ref{fig:scaling}). Over three decades of $N$, the Krylov count per step rises only from 21 to 37. The cost per step scales as $N^{1.05}$ (least-squares fit), at 11--27\,$\mu$s per DOF and step. The highest value, on M5, comes with the sparse Cholesky substitutions of \texttt{R\_ONLY} growing from 32 to 41\,\% of the solve time. The memory is governed by the near-field storage (about 3.7 times the stored $\mathbf{L}_f^{\mathrm{n+s}}$ factors plus 5.6\,kB per DOF), which depends on how the MLFMM octree subdivides the mesh rather than on $N$ alone, so a mesh with fewer DOF can need more memory than a larger one. M5 needed 309\,GB of RAM, 68\,GB of GPU memory and 26\,h for the ramp period. A workstation with 32--64\,GB of RAM and an 8--16\,GB GPU handles several $10^{5}$ DOF, and a single batch node of this type more than ten million DOF, with only the conductor meshed.

\begin{table}[!tbp]
\caption{Per-Mesh Solve Cost at 50\,Hz, Split Into the Swing-In Transient (Settling to the Periodic Regime) and the Final Benchmark Period. $\langle$GMRES/step$\rangle$ Is the Final-Period Average}\label{tab:perf}
\centering
\setlength{\tabcolsep}{4pt}
\begin{tabular}{cccccc}
\hline
\multirow{2}{*}{Mesh} & \multicolumn{2}{c}{Steps} & \multicolumn{2}{c}{Wall-clock$^{*}$} & \multirow{2}{*}{GMRES/step} \\
\cline{2-5}
 & swing-in & final & swing-in & final & \\
\hline
M1 & 477 & 109 & 2.2\,min & 0.2\,min & 19 \\
M2 & 639 & 133 & 7.6\,min & 1.0\,min & 23 \\
M3 & 597 & 121 &  22\,min & 3.6\,min & 25 \\
\hline
\multicolumn{6}{l}{\footnotesize $^{*}$Wall-clock on the node of Section~\ref{sec:verification}.}
\end{tabular}
\end{table}

\begin{figure}[!tbp]
    \centerline{\includegraphics[width=\columnwidth]{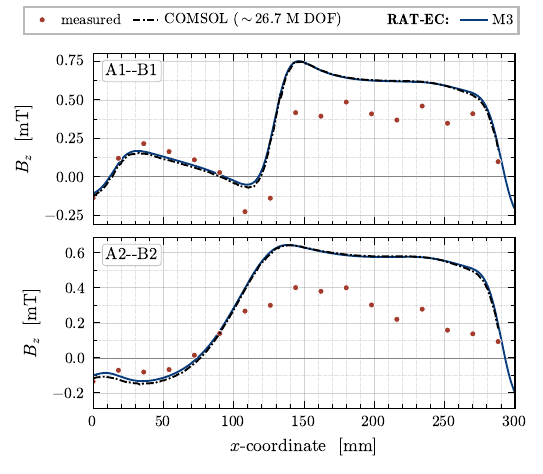}}
    \vspace{-7pt}
    \caption{Magnetic flux density $B_z$ along A1--B1 (top) and A2--B2 (bottom) at 200\,Hz and $\omega t = 90^{\circ}$: RAT-EC (blue solid; mesh M3, $\sim$175\,k DOF), the Fujiwara--Nakata reference (red markers), and a converged COMSOL FE solution ($\sim$26.7\,M DOF, black dash-dotted). The two independently converged solutions coincide but lie systematically above the measurements, for which no uncertainty was published.}
    \label{fig:b_200Hz}
\end{figure}

At $f = 200$\,Hz, the same protocol was used. At $\omega t = 0^{\circ}$, the agreement is as good as at 50\,Hz. RAT-EC on M3 matches the Fujiwara--Nakata data to $1.5\,\%$ RMS (signed peak $+4.0\,\%$) on A1--B1 and $2.4\,\%$ ($+5.9\,\%$) on A2--B2. At $\omega t = 90^{\circ}$, however, further mesh refinement did not improve the agreement with the measurements. Fig.~\ref{fig:b_200Hz} therefore compares the RAT-EC solution on M3 ($\sim$175\,k DOF) with an independently converged COMSOL solution at $\sim$26.7\,M DOF, whose $B_z$ traces were indistinguishable between $\sim$15\,M and $\sim$26.7\,M DOF in a dedicated refinement study. The two converged solutions almost coincide but lie systematically above the published data along the entire $90^{\circ}$ traces. Since the mesh-refinement studies of both codes show converged results, the discrepancy is not a numerical artefact. As the two models were built independently, the vendor's unmodified Application Gallery model and our own model built from the benchmark specification, a shared modelling error is also unlikely. The cause of the discrepancy remains unclear. Since~\cite{fujiwara1990team7} does not state the measurement uncertainty, it cannot be determined whether the simulations lie within it. This would require a new measurement at 200\,Hz, which to our knowledge has not been published.

The computational cost is higher at $200$\,Hz than at $50$\,Hz. The average $\langle$GMRES/step$\rangle$ nearly doubles, from $25$ to $42$ on M3, as the smaller $\Delta t$ increases $c_J$ and thus the weight of $c_J\mathbf{L}$ in the Jacobian. On M3, the 200\,Hz run took $1.8$\,h for a conservative integration over sixteen periods. Stopping after the eight periods required by the $0.1\,\%$ criterion would reduce this to about $55$\,min, roughly twice the $26$\,min at 50\,Hz.

\section{Conclusion and Outlook}\label{sec:conclusion}

This paper introduced RAT-EC, the first openly available volume-integral eddy-current module for unstructured meshes driven in the time domain by coils of arbitrary winding geometry. Since only the conductor is discretized, neither the air nor thin insulation gaps have to be meshed, and no outer boundary has to be treated, so problems whose narrow gaps and intricate air regions put them beyond the practical reach of standard FE analysis become tractable. The benchmarks show that this simplification does not compromise accuracy. Future work will apply RAT-EC to superconducting magnets.

Compared with closed-form solutions, the computed dissipated power converges monotonically. For the conducting sphere, the plateau error decreases to $\approx 0.6\,\%$ on the finest mesh and the error of the fitted diffusion time constant to below $0.01\,\%$, so amplitude and time scale converge together. For the conducting ring, whose current must close around a hole, the plateau error decreases to $\approx 0.2\,\%$ and the time-constant error to $\approx 2.3\,\%$ relative to the approximate thin-ring estimate, which verifies the topology-aware constraints and the $L/R$ response. For TEAM Workshop Problem~7 at 50\,Hz, the magnetic flux density along the two measurement lines agrees with the measurements within $2.4\,\%$ RMS on the finest mesh. This is at least as accurate as the converged commercial FE reference ($3.7\,\%$ and $2.5\,\%$) and is achieved with an order of magnitude fewer unknowns and only the conductor meshed. At 200\,Hz, RAT-EC and the independently built and converged FE reference agree with each other but deviate systematically from the measurements published in 1990, which suggests that this benchmark point should be re-examined.

Beyond the benchmark numbers, the results show what a volume-integral module needs in order to be both accurate and fast. Accuracy hinges on two ingredients. Firstly, the singular near and self interactions require the semi-analytical treatment, which at equal quadrature degree converges one to two orders of magnitude tighter than optimized Gauss quadrature, at lower assembly cost. Secondly, multiply connected conductors require topology-aware boundary conditions based on a cohomology basis, without which the current around a hole is suppressed entirely. As for performance, it rests on the physics-based preconditioner, which keeps the number of Krylov iterations low, and on the MLFMM, which accelerates both the current line model and the far-field inductance action. Since both are the same dense interaction problem, one multipole engine serves both. The parallel C++ implementation with GPU acceleration of the far-field and near-field products reduces the wall-clock time by a further order of magnitude.

Future extensions include (i) extending the implementation to hexahedral meshes; (ii) including the two-way circuit coupling with the coil back-EMF; (iii) coupling RAT-EC with a thermal model to capture temperature-dependent, non-linear resistivity; and (iv) coupling RAT-EC with a volume-integral non-linear magnetic material module.

All results in this paper were produced with release 1.1.0 of the open-source RAT suite~\cite{rat_group}, except M5 (release 1.2.1). The reproducibility repository~\cite{rat_validation} contains the meshes, driver and configuration files, post-processing notebooks, and Docker/HTCondor launchers needed to regenerate every figure and table. All dependencies are open-source.

\section*{Acknowledgment}
The authors used Claude (Anthropic) as a coding assistant in the development of the RAT software and to improve the text of this article. All AI-assisted content was reviewed, revised, and verified by the authors, who take full responsibility for the content of this article.

\bibliographystyle{IEEEtran}
\bibliography{rat-ec-transmag}

\end{document}